\documentclass[aps,prx,eqsecnum,onecolumn,nofootinbib]{revtex4}
\usepackage{color,graphicx,ulem}
\usepackage{amsmath,amssymb}
\usepackage{graphicx}
\usepackage{braket}
\usepackage{times,dsfont,amssymb,amsmath,amsthm,amsfonts,amsbsy,mathrsfs,bm,bbm,graphicx,graphics,epsfig,multirow,mathtools,color,bbold,empheq}

\definecolor{notas}{cmyk}{0,1,0.50,0}

\begin{document}
\title{Reframing SU(1,1) interferometry}

\author{Carlton M. Caves}
\email{ccaves@unm.edu}
\affiliation{Center for Quantum Information and Control, University of New Mexico, Albuquerque, New Mexico 87131-0001, USA}

\begin{abstract}
SU(1,1) interferometry, proposed in a classic 1986 paper by Yurke, McCall, and Klauder [Phys. Rev.~A {\bf 33}, 4033 (1986)], involves squeezing, displacing, and then unsqueezing two bosonic modes.  It has, over the past decade, been implemented in a variety of experiments.  Here I take SU(1,1) interferometry apart, to see how and why it ticks.  SU(1,1) interferometry arises naturally as the two-mode version of active-squeezing-enhanced, back-action-evading measurements aimed at detecting the phase-space displacement of a harmonic oscillator subjected to a classical force.  Truncating an SU(1,1) interferometer, by omitting the second two-mode squeezer, leaves a prototype that uses the entanglement of two-mode squeezing to detect and characterize a disturbance on one of the two modes from measurement statistics gathered from both modes.
\end{abstract}

\maketitle

\section{Introduction}
\label{sec:intro}

Quantum metrology is a field of highly sophisticated theoretical analyses~\cite{DemkowiczDobrzanski2015a}, but few applications.  The reason is that an experimenter or a technology developer is generally quite reluctant to incorporate technically difficult quantum techniques, preferring instead to push tried and true classical technologies to their limits before considering quantum enhancements.  Only when the back is against the wall are quantum techniques adopted.

The chief example of quantum metrology making a big difference is in the LIGO/VIRGO interferometric gravitational-wave detectors, where squeezed-vacuum light is being injected into the output (dark or antisymmetric) port during the current (O3) observing run~\cite{MTse2019a,Acernese2019a}.  The event rate in~O3 seems to be 5--10 times greater than in the first two observing runs (O1 and O2)~\cite{LIGOnewsreleaseAug2019}; some of this improvement can be attributed to small increases in laser power and to sensitivity enhancements in VIRGO, but at least some of it comes from employing squeezed light.  The LIGO noise data~\cite{MTse2019a} indicate an expected increase in event rate by a factor of roughly 1.5 from the use of squeezed light.

This injection of light in a squeezed-vacuum state into the dark port, proposed nearly forty years ago~\cite{Caves1981a}, reduces shot noise and is equivalent to increasing the laser power.  Not originally planned for this stage of Advanced LIGO, squeezing became nearly mandatory for sensitivity enhancements---those backs against a wall!---when Advanced LIGO fell short of the design goal for circulating power~\cite{Harry2010a,LIGO2015a} in~O1~\cite{LIGO2017d}, with only marginal power increases achieved in~O2~\cite{LIGO2017a}.  Indeed, as forecast somewhat cheekily in~\cite{Caves1981a}, when increases in power are no longer easy, ``Experimenters might then be forced to learn how to very gently squeeze the vacuum before it can contaminate the light in their interferometers.''

After initial demonstrations of shot-noise reduction in interferometers in the 1980s~\cite{MXiao1987a,Grangier1987a}, the LIGO Scientific Collaboration and the VIRGO Collaboration supported the development of single-mode squeezing technology at the audio frequencies needed in interferometric gravitational-wave detectors~\cite{Schnabel2010a,McClelland2011a}.  This multi-decade effort of technology development led to demonstrations of shot-noise reduction using squeezed light in large gravitational-wave interferometers, first in the GEO600 interferometer~\cite{LIGO2011a} and then in the LIGO Hanford interferometer~\cite{LIGO2013a} just before Initial LIGO was shut down for upgrades to Advanced \hbox{LIGO}.  This squeezing technology now provides the squeezed light for use in observing run~O3, where for the first time, squeezed light has enhanced the ability to detect actual gravitational-wave events~\cite{MTse2019a,Acernese2019a}.  Given the advanced state of techniques for generating squeezed light, it seems likely that squeezing will be a part of all future interferometric gravitational-wave detectors.

There is now another use of squeezing for enhancing an extremely challenging fundamental-physics experiment, that being the task of detecting the dark-matter axion field.  The local axions are thought to be in a highly populated, narrowband condensate that makes up the Galaxy's dark-matter halo and interacts with the electromagnetic field via a $\phi_a{\bf E}\cdot{\bf B}$ coupling~\cite{Sikivie1983a,Sikivie1985a,Krauss1985a}.  This coupling, for an electromagnetic cavity immersed in a large static magnetic field, becomes a linear interaction between the axion field and the electric field.  The axion field acts as a volume-filling classical current density that radiates into the cavity.  Detecting axions becomes a problem of linear force detection, i.e., of detecting the excitation of a cavity mode by a weak ``classical force'' that displaces the mode's complex amplitude of oscillation.

The axions are essentially at rest in the Galaxy with a velocity dispersion given by the virial velocity $v_G\simeq 200\,{\rm km/s}$ in the galactic gravitational field.  Thus the axion field excites a cavity at the rest-mass frequency, $\nu_a=m_a c^2/h=250\,{\rm MHz}\,(m_ac^2/1\,\mu{\rm eV})$, within a narrow fractional bandwith $\Delta\nu/\nu_a\simeq\frac12(v_G/c)^2\simeq 2\times 10^{-7}$, corresponding to a coherence time $\tau=1/\Delta\nu\simeq5\times 10^6\tau_a$, where $\tau_a=1/\nu_a$ is the axion period and thus the period of the cavity mode.  One can think of the axion field as exciting the cavity coherently during each time interval of duration $\tau_a$, but choosing a different phase for each such time interval.  The axion field is spatially coherent over a scale given by the de Broglie wavelength $\lambda_{\rm dB}=h/m_av_G=(c/v_G)\lambda_a\simeq1.5\times 10^3\lambda_a$, where $\lambda_a=c/\nu_a$ is the electromagnetic wavelength corresponding to frequency~$\nu_a$.

A preferred mass range for the axion, $m_a\simeq 1\,\mu{\rm eV}$--$100\,\mu{\rm eV}$, fortuitously puts the axion rest-mass frequency at microwave frequencies.  Present experiments~\cite{NDu2018a,LZhong2018a} look for excitations of a mode of a microwave cavity whose resonant frequency is progressively tuned across a wide bandwidth to search across a range of axion masses.  These cavities have ringdown times somewhat shorter than the axion coherence time.  Although squeezing can provide an advantage even in this situation~\cite{Malnou2019a}, the real advantage of squeezing comes into play when the cavities are improved to have a bandwidth much narrower than the axion bandwidth.  Then it pays to measure the displacement of the microwave mode's complex amplitude over an axion coherence time, a time over which cavity damping can be neglected and squeezed microwaves suffer little degradation.  Indeed, once in this regime, the best practical alternative for improving sensitivity---again, backs against the wall---is to use back-action-evading measurements of a single quadrature displacement of the cavity mode~\cite{Thorne1978a,Hollenhorst1979a,Unruh1979a,Thorne1979a,Caves1980b,Braginsky1980a,Caves1983a,Bocko1996a} or, equivalently, to use active squeezing of that quadrature to achieve the same effect~\cite{Burd2019a}.  The experiment becomes a version of a back-action-evading (or quantum nondemolition) measurement of the displacement of one quadrature component of a harmonic oscillator, in this case, a microwave-cavity~mode.

To measure both quadratures of the axion-induced displacement, one can use two cavities that are within a spatial coherence length of one another, or one can use an equivalent scheme in which only one of the cavities is subjected to the axion displacement, but the two cavities are coupled by putting them in a two-mode squeezed state.  This latter idea, proposed in the axion context by~\cite{HZheng2016au}, is equivalent to using an SU(1,1) interferometer of the sort introduced in a classic 1986 paper by Yurke, McCall, and Klauder~\cite{Yurke1986a}.  That, at last, brings us around to the subject of this paper: to examine SU(1,1) interferometry and the prominent role it and variants of it play in quantum metrology and quantum information.

Over the last decade, a number of experimental groups have implemented SU(1,1) interferometry in a variety of forms~\cite{JJing2011a,Hudelist2014a,BChen2015a,Linnemann2016a,BAnderson2017a,BAnderson2017b,Manceau2017a,JLi2018a,PGupta2018a,SLiu2018a,YLiu2019a,Frascella2019a}. A brief review of SU(1,1) interferometry within the context of other contemporary uses of squeezed light, which touches on some of questions considered in this paper, can be found in~\cite{Lawrie2019a}.

This paper reviews, in Sec.~\ref{sec:SU11}, the ideas behind SU(1,1) interferometry and formulates two distinct scenarios for its use.  The first scenario involves detecting \textit{itinerant signals\/} that are applied to \textit{persistent modes}, by which I mean that the modes must be checked repeatedly to provide evidence of a signal that acts only occasionally or with varying amplitude and phase.  The second scenario, involves a \textit{persistent signal} (or disturbance) that can be consulted by \textit{itinerant modes\/} that successively probe the signal.

The first scenario, examined in Sec.~\ref{sec:BAE}, is the province of high-resolution, back-action-evading measurements of a quadrature displacement produced by a ``classical force'' acting on a persistent mode.  The discussion in Sec.~\ref{sec:BAEbothquadratures} leads up to showing that an SU(1,1) interferometer is a squeezing-enhanced version of back-action-evading measurements for detecting both (orthogonal) quadrature displacements, with the enhancement coming from the fact that two-mode squeezing is noiseless, phase-sensitive amplification and de-amplification of Einstein-Podolsky-Rosen (EPR) variables of the two modes.

In the second scenario, which occupies Sec.~\ref{sec:PerSigItMode}, the goal is not high sensitivity in a single shot, but rather reliable detection or characterization of a persistent disturbance over many trials by taking advantage of the modal entanglement within an SU(1,1) interferometer.  This scenario is illustrated by an example from my own recent work, a protocol~\cite{RahimiKeshari2019au} for characterizing a lossy, passive linear optical network in randomized boson sampling~\cite{Lund2014a}.

The bottom line is that SU(1,1) interferometry is not really about interferometry at all.  In the first scenario, the use of SU(1,1) operations is all about high-resolution linear force detection using the noiseless amplification and de-amplification of squeezers as the primary resource.  The prominence of noiseless linear amplification and de-amplification and measurement of linear observables prompts a renaming of SU(1,1) interferometers as SU(1,1) displacement detectors.  The second scenario is all about reliable detection and/or characterization of a disturbance on one mode using the entanglement introduced by a two-mode squeezer as the primary resource.  In this second scenario, the omission of the second two-mode squeezer severs any connection with interferometry.

Much, perhaps all, of what is discussed in this paper is not new.  The purpose of the paper is to bring together a wide variety of SU(1,1)-based measurement techniques, so that one can easily see the connections among and distinctions between them.

\section{SU(1,1) interferometry}
\label{sec:SU11}

\begin{figure}
\centering
\includegraphics[width=0.7\columnwidth]{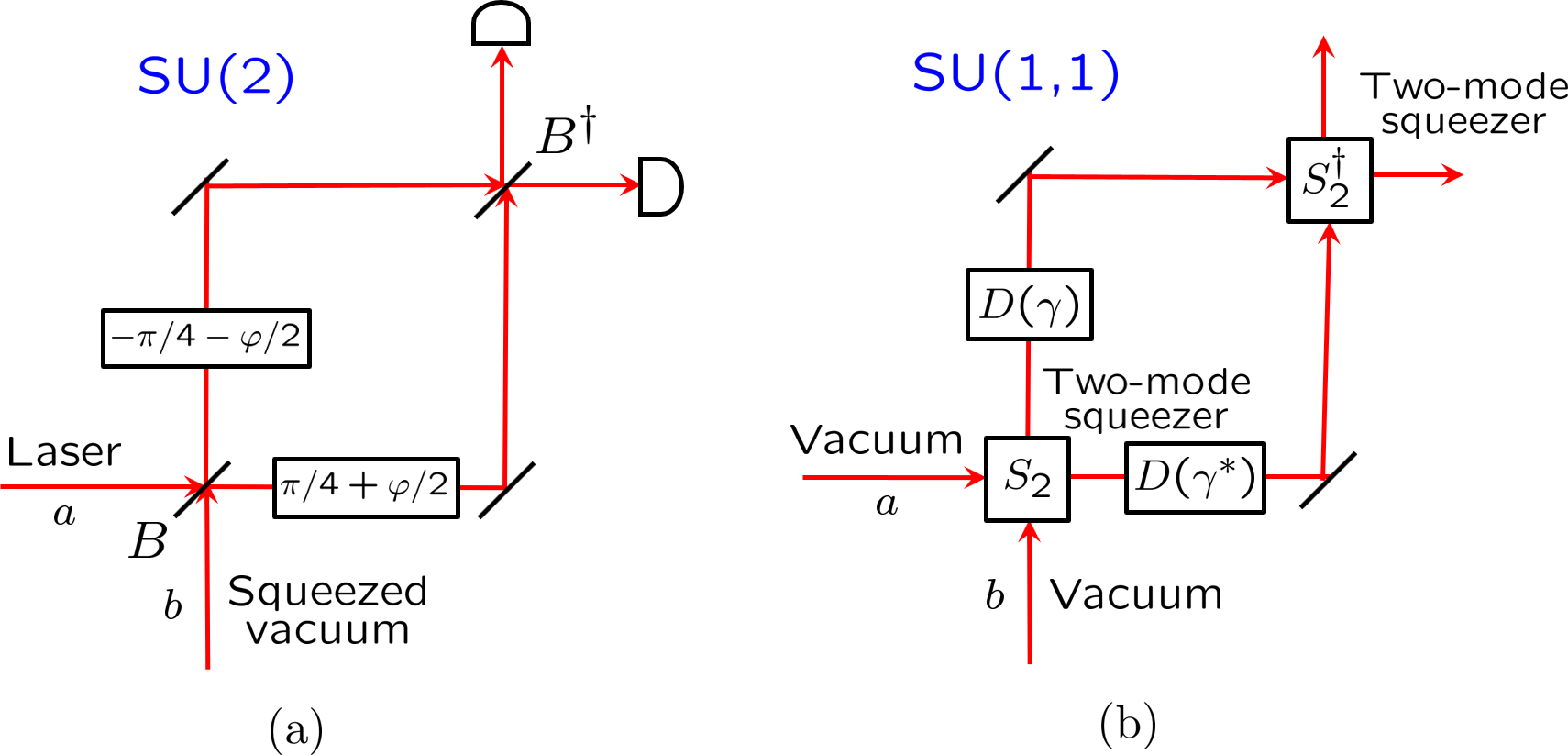}
\caption{(a)~Mach-Zehnder interferometer as a prototype for SU(2) interferometry.  The operations in the interferometer are generated by the Schwinger operators $J_1$, $J_2$, and $J_3$ of Eq.~(\ref{eq:Js}), which obey the angular-momentum commutation relations~(\ref{eq:Jcomm}) and thus generate the group SU(2).  The interferometer is powered by a laser that prepares a coherent state in the (horizontal) input mode~$a$; the overall phase reference for the interferometer is the phase of this coherent state.  The laser light is split between two arms by the initial 50-50 beamsplitter $B=e^{(\hat a\hat b^\dagger-\hat a^\dagger\hat b)\pi/4}=e^{-iJ_2\pi/2}$ and is then subjected to equal and opposite phase shifts in the two arms, $e^{i(\hat a^\dagger\hat a-\hat b^\dagger\hat b)(\pi/4+\varphi/2)}=e^{iJ_3\pi/2}e^{iJ_3\varphi}$.  The beams are recombined at a second 50-50 beamsplitter $B^\dagger$, which converts the differential phase shift $\varphi$ to amplitude changes in the output beams; these are detected at two photodetectors, the photocounts are differenced, and thus the detection is a measurement of $\hat a^\dagger\hat a-\hat b^\dagger\hat b=2J_3$.  The differential phase shift $e^{iJ_3\pi/2}$ is included so that the interferometer operates, when $\varphi=0$, with equal amounts of laser light exiting the two output ports; it could be included in either the input or the output beamsplitter.  The overall transformation through the interferometer, $e^{iJ_2\pi/2}e^{iJ_3\pi/2}e^{iJ_3\varphi}e^{-iJ_2\pi/2}=e^{-iJ_1\pi/2}e^{-iJ_1\varphi}$, is a beamsplitter whose transmission and reflection are changed from 50-50 by the signal $\varphi$.  To reduce the shot noise that limits measuring $\varphi$, one injects squeezed vacuum into the (vertical) input port~$b$ (the dark or antisymmetric port), with the reduced-noise (squeezed) quadrature being the one that is out of phase with the laser light after passage through the initial beamsplitter. (b)~SU(1,1) interferometer.  The SU(2) generator $J_2$ is replaced by the SU(1,1) generator $K_2$, one of the three operators of Eq.~(\ref{eq:Ks}) that generate the group SU(1,1); to be precise, the beamsplitters of the Mach-Zehnder interferometer are replaced by two-mode squeezers $S_2(r)=e^{r(\hat a\hat b-\hat a^\dagger \hat b^\dagger)}=e^{-2iK_2 r}$.  The squeezers being active elements, the interferometer doesn't need to be powered, so can have vacuum inputs.  The overall phase reference is set by the phase of whatever pump beam drives the two-mode squeezers.  After the first squeezer, the signal is applied: it consists of phase-space displacements, $D(\gamma^*)=e^{\gamma^*\hat a^\dagger-\gamma\hat a}$ on mode~$a$ and $D(\gamma)=e^{\gamma\hat b^\dagger-\gamma^*\hat b}$ on mode~$b$.  The second two-mode squeezer acts oppositely to the first; in the absence of the signal, the modes are restored to vacuum.  The signal is detected by measuring two EPR variables of the output mode, as is discussed in Sec.~\ref{sec:BAEbothquadratures}.}
\label{fig:interferometry}
\end{figure}

Yurke, McCall, and Klauder~\cite{Yurke1986a} introduced the notion of SU(1,1) interferometry by replacing the beamsplitters of a standard SU(2) interferometer with active elements now called two-mode squeezers.

A standard interferometer uses beamsplitters acting on a pair of modes, $a$ and $b$, to send waves down two different paths and to bring those waves into interference after they have received phase shifts that one wants to detect.  A standard interferometer is called an SU(2) interferometer because the operations on the two modes are generated by the three Schwinger operators,
\begin{align}\label{eq:Js}
\begin{split}
J_1&=J_x=\frac12(\hat a^\dagger\hat b+\hat a\hat b^\dagger)\,,\\
J_2&=J_y=-\frac12 i(\hat a^\dagger\hat b-\hat a\hat b^\dagger)\,,\\
J_3&=J_z=\frac12(\hat a^\dagger\hat a-\hat b^\dagger\hat b)\,,
\end{split}
\end{align}
which obey the angular-momentum commutation relations,
\begin{align}\label{eq:Jcomm}
[J_j,J_k]=iJ_l\epsilon_{ljk}\,,
\end{align}
and thus generate the group SU(2).

Yurke, McCall, and Klauder suggested replacing the beamsplitters of a standard interferometer with active elements, two-mode squeezers, which are described by the unitary operator,
\begin{align}\label{eq:S2}
S_2(r)=e^{r(\hat a\hat b-\hat a^\dagger\hat b^\dagger)}=e^{-2iK_2 r}\,,
\end{align}
where $r$ is called the squeeze parameter.  The generator $K_2$ is one of a set of three operators,
\begin{align}\label{eq:Ks}
\begin{split}
K_0=K_t&=\frac12(\hat a^\dagger\hat a+\hat b^\dagger\hat b+1)=\frac12(\hat a^\dagger\hat a+\hat b\hat b^\dagger)\,,\\
K_1=K_x&=\frac12(\hat a\hat b+\hat a^\dagger \hat b^\dagger)\,,\\
K_2=K_y&=\frac12 i(\hat a\hat b-\hat a^\dagger \hat b^\dagger)\,,
\end{split}
\end{align}
which have the commutators of the group~SU(1,1),
\begin{align}\label{eq:Kcomm}
[K_\alpha,K_\beta]=iK^\gamma\epsilon_{\gamma\alpha\beta}=iK_\gamma{\epsilon^\gamma}_{\alpha\beta}\,.
\end{align}
Here indices are raised and lowered using the Minkowski metric for two spatial dimensions, $||\eta_{\alpha\beta}||={\rm diag}(-1,1,1)$, so $K^0=\eta^{0\alpha}K_\alpha=-K_0$.

This is a good place for the reader to consult Fig.~\ref{fig:interferometry}, which summarizes the transition from SU(2) to SU(1,1) interferometry in language like that used by Yurke, McCall, and Klauder.  In considering applications of SU(1,1) interferometry in quantum metrology and, more generally, in quantum information science, it helps, right at the start, to distinguish two different scenarios.

The first scenario involves detecting \textit{itinerant signals\/} that are applied to \textit{persistent modes\/}: the modes persist in the laboratory and must be checked repeatedly to provide evidence of a signal that acts occasionally and unpredictably.  In this situation, one can think in terms of a sequence of time intervals, during each of which the modes begin in an appropriate quantum state and are measured at the end to determine whether a signal has acted.\footnote{Often, what is actually performed, especially for microwave modes, is a continuous measurement with a bandwidth given by the inverse of the duration of the time intervals, but it is easier to think in terms of and to draw quantum circuits for a sequence of time intervals.}  The crucial point is that \textit{the measurement itself prepares the persistent modes in an appropriate quantum state, ready for the next time interval and the next measurement\/}; this is the defining feature of back-action-evading measurements.  To visualize this scenario in the SU(1,1) interferometer of Fig.~\ref{fig:interferometry}(b), one should imagine the measured modes at the output of the interferometer circling around to become the input for the next round.

This first scenario is thus about subvacuum-noise, back-action-evading measurements of a quadrature component of a persistent mode to detect a ``classical force'' that, acting linearly on the mode, displaces its complex amplitude of oscillation.  Proposed for detecting the action of a gravitational wave on a mode of oscillation of a several-ton metallic bar, such back-action-evading measurements are equally suited to metrological applications of the wide range of high-$Q$ mechanical oscillators being developed in optomechanics~\cite{Kippenberg2008a,Meystre2013a,Aspelmeyer2014a} and to axion detection that uses a persistent mode of a microwave resonator.

Subvacuum-noise quadrature measurements prepare the persistent mode(s) in a squeezed state, with no active squeezing required.  Active squeezing gets into the picture when the available quadrature measurements do not have subvacuum-noise resolution.  In that situation, active squeezing achieves subvacuum-noise resolution by using its noiseless, phase-sensitive amplification and de-amplification of quadrature components~\cite{Caves1982a} to make a quadrature displacement stand out above whatever noise governs the quadrature measurement~\cite{Burd2019a}.  These possibilities are explored in detail in Sec.~\ref{sec:BAE}.

The second scenario for SU(1,1) interferometry involves a \textit{persistent signal} (or disturbance) that can be consulted over and over again by a succession of \textit{itinerant modes}.  The goal in this second scenario, which is the subject of Sec.~\ref{sec:PerSigItMode}, is not high sensitivity in a single shot, but rather the ability over many trials to detect reliably and/or to characterize a disturbance.  In this scenario, one should think of the itinerant modes as being demolished on detection, with new modes injected in each round to probe the disturbance.  One can omit as irrelevant the second two-mode squeezer in the SU(1,1) interferometer of Fig.~\ref{fig:interferometry}, thus reducing the device to what is called a truncated SU(1,1) interferometer~\cite{BAnderson2017a,BAnderson2017b,PGupta2018a}.   The key feature of these protocols is that one seeks to detect over many shots a disturbance to one of the modes by making measurements on the other mode or on both modes, with the entangled correlations introduced by the two-mode squeezer making this possible.

\section{Back-action-evading measurements: Itinerant signal, persistent modes}
\label{sec:BAE}

\subsection{BAE meaurements of quadrature displacement}
\label{sec:BAEsinglequadrature}

Focus now on measurements of a quadrature displacements of a persistent mode $a$, as in bar gravitational-wave detection or axion detection, and write the modal annilation operator in terms of position and momentum quadrature components:
\begin{align}
\hat a=\frac{1}{\sqrt2}(\hat x+i\hat p)\,.
\end{align}
A subtle, but important point is that all these operators are constants in the absence of the signal one wishes to detect; i.e., one has gone to a rotating frame in which the free, harmonic evolution at the modal frequency $\omega$ is removed from $\hat a$, $\hat x$, and $\hat p$.  When one talks about measuring quadrature components or about measuring the complex amplitude of a mode, one is talking about measuring these constant quantities.  This is different from measuring the position or momentum, which co-evolve harmonically at the modal frequency.

\begin{figure}
\centering
\includegraphics[width=0.8\columnwidth]{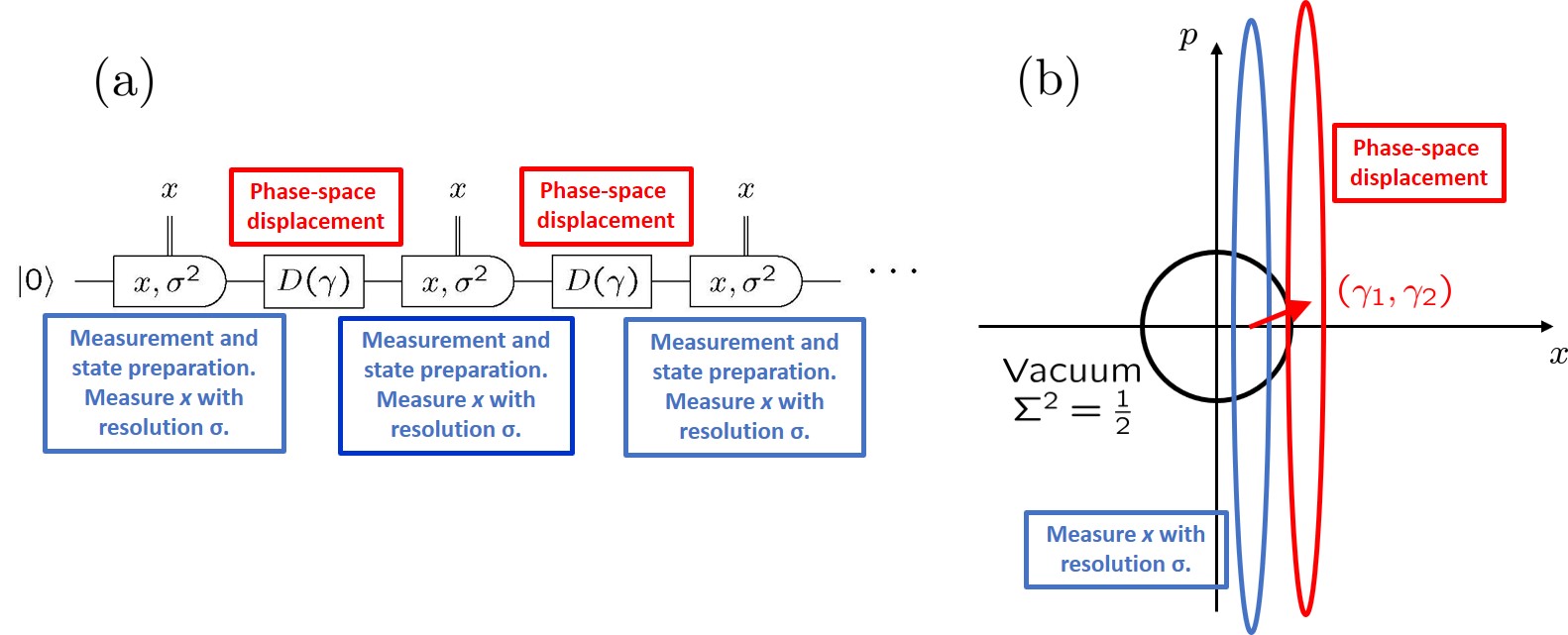}
\caption{(a)~Sequence of BAE position-quadrature measurements of a persistent mode.  The mode begins, say, in vacuum, as shown here, and is subjected to a sequence of position-quadrature ($x$) measurements, each with resolution $\sigma<1/\sqrt2$.  Successive pairs of measurements detect the $x$ component of a phase-space displacement, represented by the displacement operator $D(\gamma)$ of Eq.~(\ref{eq:Dgamma}), acting on the mode between the two measurements.  In the language of quantum circuits, the essence of back-action-evading measurements is that there is a persistent mode, represented by a quantum wire that persists through the measurements; each measurement, whose outcome $x$ is represented by the vertical, double (classical) wire, prepares the mode in a squeezed state, ready for the next measurement to reveal an intervening phase-space displacement.  (b)~Phase-space depiction of the initial measurement in~(a).  The mode, initially in vacuum, represented by the black, circular noise ellipse, is subjected to a position-quadrature measurement, whose outcome~$x$ is chosen from a Gaussian distribution with variance $\sigma^2+\Sigma_0^2=\sigma^2+\frac12$.  After the measurement, the mode is left in a squeezed state, represented by the blue noise ellipse, which has position-quadrature variance $\alt\sigma^2$.  The complex phase-space displacement $\gamma=(\gamma_1+i\gamma_2)/\sqrt2$ moves the mode to the red noise ellipse.  A subsequent position-quadrature measurement detects $\gamma_1$ with resolution $\simeq\sigma$.  As the protocol proceeds, the modal state becomes highly squeezed, approaching a position-quadrature eigenstate; it is then easy to see that successive pairs of measurements determine the intervening position-quadrature displacement with resolution $\sqrt2\sigma$.  The analysis in the text shows that this $\sqrt2\sigma$ resolution holds for all pairs of successive measurements.}
\label{fig:BAE}
\end{figure}

Measuring the constant quadrature components is the crucial feature of back-action-evading (BAE)    measurements~\cite{Thorne1978a,Hollenhorst1979a,Unruh1979a,Thorne1979a,Caves1980b,Braginsky1980a,Caves1983a,Bocko1996a}.  When measuring modes of the electromagnetic field, at microwave or optical frequencies, the natural measurements are measurements of these constant quantities: homodyne measurement is a measurement of a quadrature component, and heterodyne measurement reports a mode's constant complex amplitude.  In contrast, for mechanical oscillators, it is natural to think in terms of coupling to the position or momentum, not to quadrature components; in this situation, measuring a quadrature component involves modulating the coupling at the oscillator frequency and averaging over several oscillator periods.

Ideas of subvacuum measurement resolution have been mainly used on electromagnetic modes, where homodyne and heterodyne, the natural measurements, measure conserved quantities.  This has served to obscure the distinctive feature of BAE measurements for some~\cite{Monroe2011a}.  The point of back-action-evading (or quantum nondemolition) measurements is not that the measurement is projective or that the measured system is left untouched except for the measurement back action, but rather that the persistent system in the state left by measurement is ready to be used again in a protocol of repeated measurements that evade the noise left by the back action of previous measurements.  This generally means, as here, measuring a conserved quantity, but for a pair of modes, it can mean measuring harmonically evolving observables in what is called a quantum-mechanics-free subsystem~\cite{MTsang2012a}.

I assume throughout that the measurements of quadrature components dominate the persistent mode's other couplings to the external world, this being the regime where subvacuum-noise resolution can be achieved.  In practice, this means that the measurements dominate the effects of finite temperature and dissipation, which are therefore ignored.

Consider now a sequence of measurements of the position-quadrature component.  The measurements are described by Kraus operators,
\begin{align}\label{eq:Kxsigma2}
\sqrt{dx}\,K_{x,\sigma^2}
=\sqrt{dx}\int_{-\infty}^\infty dy\,\frac{e^{-(y-x)^2/4\sigma^2}}{(2\pi\sigma^2)^{1/4}}|y\rangle\langle y|\,,
\end{align}
where $|y\rangle$ are eigenstates of the position quadrature $\hat x$, $x$ denotes the outcome of the measurement, and $\sigma$ is the measurement resolution.  I refer to measurements described by these Kraus operators as BAE homodyne measurements, in this case, of the position quadrature.  These Kraus operators are Hermitian, and the POVM elements are given by
\begin{align}
dx\,E_{x,\sigma^2}
=dx\,K_{x,\sigma^2}^\dagger K_{x,\sigma^2}=dx\int_{-\infty}^\infty dy\,\frac{e^{-(y-x)^2/2\sigma^2}}{\sqrt{2\pi\sigma^2}}|y\rangle\langle y|\,.
\end{align}
Vacuum resolution, which I here call, following~\cite{Bachor2004a}, the \textit{quantum noise limit\/} (QNL), is set by the size of zero-point fluctuations and is thus given by $\sigma^2=1/2$.  I assume here that the quadrature measurements have resolution better than the QNL, i.e., $\sigma^2<1/2$.

Between measurements, a force might act on the system, its effect on the mode described by a displacement operator,
\begin{align}\label{eq:Dgamma}
D(\gamma)=e^{\gamma \hat a^\dagger-\gamma^*\hat a}=e^{i(\gamma_2\hat x-\gamma_1\hat p)}\,,\qquad\gamma=\frac{1}{\sqrt2}(\gamma_1+i\gamma_2)\,.
\end{align}
The phase-space displacement of the complex amplitude, $\gamma=(\gamma_1+i\gamma_)/\sqrt2$, divides into position and momentum quadrature displacements, $\gamma_1$ and $\gamma_2$.  Figure~\ref{fig:BAE} translates the sequence of measurements into a quantum circuit and depicts graphically how the initial measurement on a vacuum input leaves the system in a squeezed state that can be used to detect a subsequent quadrature displacement with sensitivity $\sim\sigma$.

I sketch here the analysis of a sequence of BAE homodyne measurements.  If the persistent mode begins in a pure Gaussian state, such as vacuum, it remains in a pure Gaussian state throughout the entire sequence.  After $n$ measurements, the system is in state $|\psi_n\rangle$, with a Gaussian wave function whose position-quadrature mean and variance are~$\langle\hat x\rangle_n$ and $\Sigma_n^2$.  The next phase-space displacement $D(\gamma_{n+1})$ changes the mean value to $\langle\hat x\rangle_n+\gamma_{1,n+1}$; the position-quadrature probability distribution is
\begin{align}
\big|\langle y|D(\gamma_{n+1})|\psi_n\rangle\big|^2
=\frac{1}{\sqrt{2\pi\Sigma_n^2}}\exp\!\bigg({-}\frac{\big(y-\langle\hat x\rangle_n-\gamma_{1,n+1}\big)^2}{2\Sigma_n^2}\bigg)\,.
\end{align}
The outcome $x_{n+1}$ of the $(n+1)$th measurement is drawn from a Gaussian distribution,
\begin{align}
p(x_{n+1})
=\langle\psi_n|D^\dagger(\gamma_{n+1})E_{x_{n+1},\sigma^2}D(\gamma_{n+1})|\psi_n\rangle
=\int_{-\infty}^\infty dy\frac{e^{-(y-x_{n+1})^2/2\sigma^2}}{\sqrt{2\pi\sigma^2}}\big|\langle y|D(\gamma_{n+1})|\psi_n\rangle\big|^2\,,
\end{align}
which has mean $\langle x_{n+1}\rangle=\langle\hat x\rangle_n+\gamma_{1,n+1}$ and variance $\sigma^2+\Sigma_n^2$; one can write the outcome as
\begin{align}\label{eq:xnplus1}
x_{n+1}=\langle\hat x\rangle_n+\gamma_{1,n+1}+\sqrt{\sigma^2+\Sigma_n^2}\,W_{n+1}\,,
\end{align}
where $W_{n+1}$ is a zero-mean, unit-variance Gaussian random variable.  These random variables are independent from one measurement to the next.

The modal state after the $(n+1)$th measurement,
\begin{align}
|\psi_{n+1}\rangle=\frac{K_{x_{n+1},\sigma^2}D(\gamma_{n+1})|\psi_n\rangle}{\sqrt{p(x_{n+1})}}\,,
\end{align}
has Gaussian wave function,
\begin{align}
\langle y|\psi_{n+1}\rangle
=\frac{e^{-(y-x_{n+1})^2/4\sigma^2}}{(2\pi\sigma^2)^{1/4}}
\frac{\langle y|D(\gamma_{n+1})|\psi_n\rangle}{\sqrt{p(x_{n+1})}}\,,
\end{align}
and thus position-quadrature probability distribution,
\begin{align}
\big|\langle y|\psi_{n+1}\rangle\big|^2
=\frac{e^{-(y-x_{n+1})^2/2\sigma^2}\big|\langle y|D(\gamma_{n+1})|\psi_n\rangle\big|^2}{p(x_{n+1})\sqrt{2\pi\sigma^2}}\,.
\end{align}
This distribution, as the product of two Gaussians, has position-quadrature variance,
\begin{align}
\Sigma_{n+1}^2=\sigma^2\frac{\Sigma_n^2}{\sigma^2+\Sigma_n^2}\,,
\end{align}
and position-quadrature mean,
\begin{align}\label{eq:meanx}
\begin{split}
\langle\hat x\rangle_{n+1}
&=\frac{\Sigma_n^2}{\sigma^2+\Sigma_n^2}x_{n+1}+\frac{\sigma^2}{\sigma^2+\Sigma_n^2}\big(\langle\hat x\rangle_n+\gamma_{1,n+1}\big)\\
&=\langle\hat x\rangle_n+\gamma_{1,n+1}+\frac{\Sigma_n^2}{\sqrt{\sigma^2+\Sigma_n^2}}W_{n+1}\,.
\end{split}
\end{align}

The iterative equations for the variance are easily solved, giving
\begin{align}
\Sigma_n^2&=\frac{\sigma^2}{n+\sigma^2/\Sigma_0^2}\,.
\end{align}
After just one measurement, the variance is reduced to below $\sigma^2$, and after many measurements, the state becomes highly squeezed, ultimately approaching a position-quadrature eigenstate.

One estimates a quadrature displacement as the difference of successive measurement outcomes, i.e.,
\begin{align}\label{eq:gamma1n}
\begin{split}
\gamma_{1,n}^{\rm est}=x_{n}-x_{n-1}
&=\gamma_{1,n}+\sigma W_n\sqrt{1+\Sigma_{n-1}^2/\sigma^2}-\sigma W_{n-1}\frac{1}{\sqrt{1+\Sigma_{n-2}^2/\sigma^2}}\\
&=\gamma_{1,n}+\sigma W_n\sqrt{\frac{n+\sigma^2/\Sigma_0^2}{n-1+\sigma^2/\Sigma_0^2}}
-\sigma W_{n-1}\sqrt{\frac{n-2+\sigma^2/\Sigma_0^2}{n-1+\sigma^2/\Sigma_0^2}}\,.
\end{split}
\end{align}
The estimator is unbiased, $\langle\gamma_{1,n}^{\rm est}\rangle=\gamma_{1,n}$, and the resolution for measuring the quadrature displacement is given by the square root of the estimator variance,
\begin{align}\label{eq:gamma1nest}
\big\langle(\Delta\gamma_{1,n}^{\rm est})^2\big\rangle^{1/2}=\sqrt2\sigma\,,
\end{align}
and is independent of $n$.

Although this analysis arises naturally from the description of a sequence of BAE homodyne measurements, by tracking both the outcomes and the mean values of $\hat x$ in terms of the Gaussian random variables $W_n$, it mixes modal and measurement variances and thus obscures the reason for the $\sqrt2$ in the ultimate resolution~(\ref{eq:gamma1nest}).  To gain understanding, iterate Eq.~(\ref{eq:meanx}) to obtain
\begin{align}
\langle\hat x\rangle_n=\sum_{j=1}^n\gamma_{1,j}+\sum_{j=1}^n\frac{\Sigma_{j-1}^2}{\sqrt{\sigma^2+\Sigma_{j-1}^2}}W_j\,,
\end{align}
where we assume $\langle\hat x\rangle_0=0$ without loss of generality, and then plug this into Eq.~(\ref{eq:xnplus1}) to find
\begin{align}
x_{n+1}=\sum_{j=1}^{n+1}\gamma_{1,j}+Z_{n+1}\,,
\end{align}
where the Gaussian random variable $Z_{n+1}$ is defined by
\begin{align}
Z_{n+1}=\sqrt{\sigma^2+\Sigma_n^2}\,W_{n+1}+\sum_{j=1}^n\frac{\Sigma_{j-1}^2}{\sqrt{\sigma^2+\Sigma_{j-1}^2}}W_j\,.
\end{align}
The correlation between these variables is, for $n\ge m$,
\begin{align}\label{eq:Znmplus1}
\langle Z_{n+1}Z_{m+1}\rangle
=\sigma^2\delta_{nm}+\Sigma_m^2+\sum_{j=1}^m\frac{\Sigma_{j-1}^4}{\sigma^2+\Sigma_{j-1}^2}
=\sigma^2\delta_{nm}+I_m\,.
\end{align}
Notice now that
\begin{align}
I_m-I_{m-1}
=\Sigma_m^2-\Sigma_{m-1}^2+\frac{\Sigma_{m-1}^4}{\sigma^2+\Sigma_{m-1}^2}
=\Sigma_m^2-\frac{\sigma^2\Sigma_{m-1}^2}{\sigma^2+\Sigma_{m-1}^2}=0\,,
\end{align}
which implies that $I_m=I_0=\Sigma_0^2$.  The correlations~(\ref{eq:Znmplus1}) thus simplify to
\begin{align}
\langle Z_{n+1}Z_{m+1}\rangle=\sigma^2\delta_{nm}+\Sigma_0^2\,,
\end{align}
and this allows us to relate $Z_{n+1}$ to new, uncorrelated, unit-variance Gaussian random variables, $W'_{n+1}$ and $W'_0$, i.e.,
\begin{align}
Z_{n+1}=\sigma W'_{n+1}+\Sigma_0 W'_0\,,
\end{align}
which leads to
\begin{align}\label{eq:outcomesimple}
x_{n+1}=\sum_{j=1}^{n+1}\gamma_{1,j}+\sigma W'_{n+1}+\Sigma_0 W'_0\,,
\end{align}
This result has the obvious interpretation that the outcome of a measurement has mean given by all the displacements up to that point and variance given by the sum of the variance of the initial wave function and the resolution of the measurement.\footnote{An astute reader will recognize that Eq.~(\ref{eq:outcomesimple}) is a transcription of the meter model that gives the Kraus operators~(\ref{eq:Kxsigma2}).}  The estimator~(\ref{eq:gamma1n}) now takes the form $\gamma_{1,n}^{\rm est}=\gamma_{1.n}+\sigma W'_n-\sigma W'_{n-1}$, clearly showing how the factor of $\sqrt2$ in the estimator variance~(\ref{eq:gamma1nest}) arises from equal contributions from the two successive measurements required to estimate an intervening displacement.

\begin{figure}
\centering
\includegraphics[width=0.9\columnwidth]{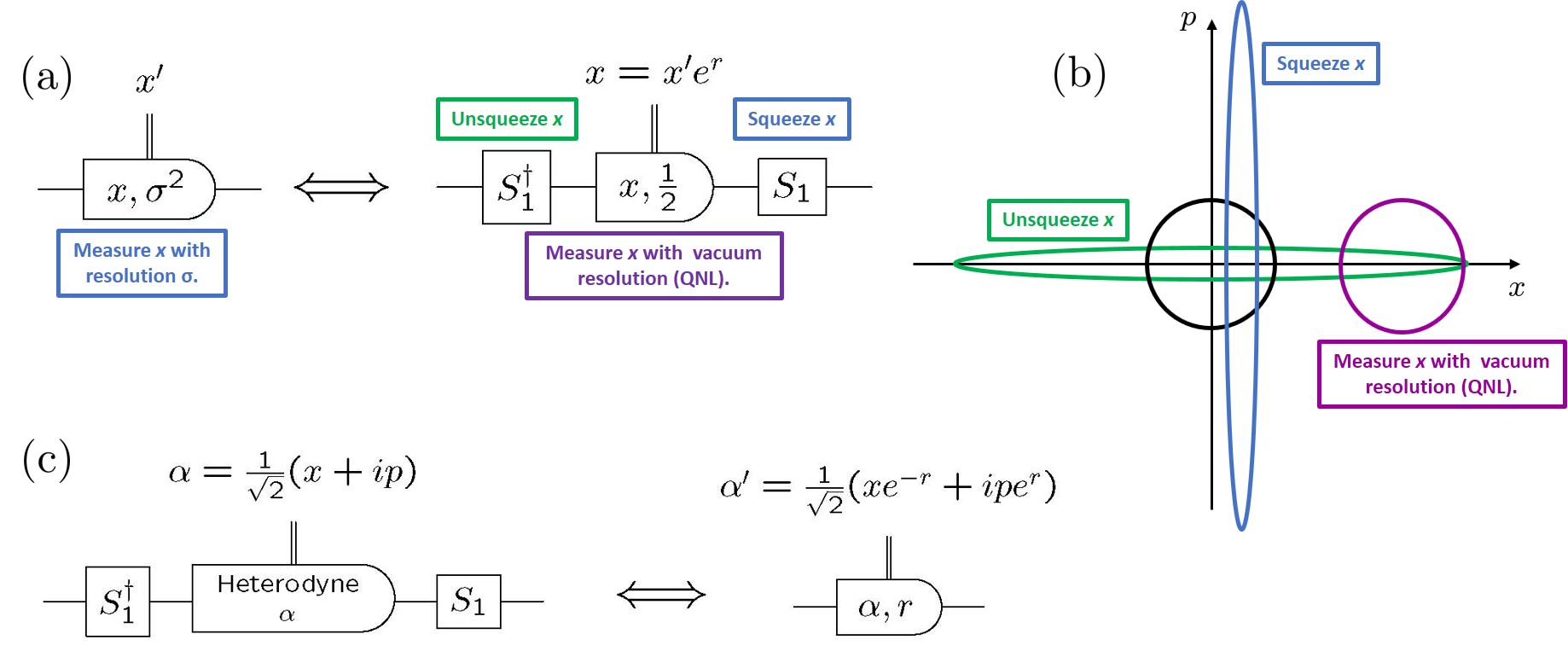}
\caption{(a)~BAE homodyne measurement of the position quadrature~$x$ with resolution $\sigma=e^{-r}/\sqrt2$ is equivalent to conjugating with squeeze operators a measurement of $x$ with vacuum resolution.  This substitutes active squeezing for measurements with subvacuum-noise resolution.  (b)~Phase-space depiction of the active squeezing circuit in~(a).  Suppose the mode begins in vacuum, represented by the black, circular noise ellipse.  Unsqueezing $x$ ($S_1^\dagger$) produces the green noise ellipse.  A measurement at the QNL, whose outcome~$x$ is chosen from a Gaussian distribution with variance $\frac12+\frac12 e^{2r}=\frac12+1/4\sigma^2$, leaves the mode in a state with the purple noise ellipse.  Squeezing $x$ ($S_1$) then leaves the mode with the blue noise ellipse, identical to the blue noise ellipse after a BAE measurement on vacuum with resolution $\sigma$ in Fig.~\ref{fig:BAE}(b).  (c)~Heterodyne measurement, with outcome $\alpha=(x+ip)/\sqrt2$, can be used in place of BAE homodyne measurement to achieve sub-QNL resolution.  The left circuit conjugates a heterodyne measurement, which projects onto coherent states $|\alpha\rangle$, with squeeze operators; the right circuit is the equivalent measurement that projects onto squeezed states $|\alpha',r\rangle=S_1|\alpha\rangle$.  The analysis in the text shows that these measurements achieve the same resolution, $\sqrt2\sigma$, as BAE homodyne measurements, with the advantage that the weak determination of the $p$ quadrature keeps the modal state from becoming highly squeezed as the protocol proceeds.}
\label{fig:BAEtoSqueeze}
\end{figure}

BAE homodyne measurements are by themselves sufficient for sub-QNL detection of a classical force---no active squeezing required---provided the measurements are done at a subvacuum noise level.  Active squeezing comes into play if one doesn't have the ability to do such sensitive measurements; then active squeezing can substitute for more sensitive measurements.  To see this, let's first get a little notation straight, introducing the single-mode squeeze operator,
\begin{align}\label{eq:S1}
S_1(r)=e^{r(\hat a^2-\hat a^{\dagger 2})/2}=e^{ir(\hat x\hat p+\hat p\hat x)/2}\,,
\end{align}
which transforms the quadrature components according to
\begin{align}
S_1^\dagger\hat x S_1=\hat x\,e^{-r}\,,\qquad
S_1^\dagger\hat p S_1=\hat p\,e^r\,.
\end{align}
The transformation of the position quadrature implies that $S_1|y\rangle=e^{-r/2}|ye^{-r}\rangle$.  Consider now how the single-mode squeeze operator transforms the BAE homodyne Kraus operators~(\ref{eq:Kxsigma2}),
\begin{align}
\begin{split}
S_1\,\sqrt{dx}\,K_{x,1/2}S_1^\dagger
&=\sqrt{dx}\int dy\,\frac{e^{-(y-x)^2/2}}{\pi^{1/4}}S_1|y\rangle\langle y|S_1^\dagger\\
&=\sqrt{dx}\int d(ye^{-r})\,\frac{e^{-(ye^{-r}-xe^{-r})^2/2e^{-2r}}}{\pi^{1/4}}|ye^{-r}\rangle\langle ye^{-r}|\\
&=\sqrt{d(xe^{-r})}\int dy\,\frac{e^{-(y-xe^{-r})^2/4\sigma^2}}{(2\pi\sigma^2)^{1/4}}|y\rangle\langle y|\\
&=\vphantom{\int}\sqrt{dx'}\,K_{x',\sigma^2}\,,
\end{split}
\end{align}
where $\sigma^2=\frac12 e^{-2r}$ and $x'=xe^{-r}$.  Conjugating a position-quadrature measurement at the QNL with squeeze operators transforms it to a measurement with resolution $\sigma=e^{-r}/\sqrt2$.  For a better understanding of what this means and how it works, consult Fig.~\ref{fig:BAEtoSqueeze}.

\begin{figure}
\centering
\includegraphics[width=0.8\columnwidth]{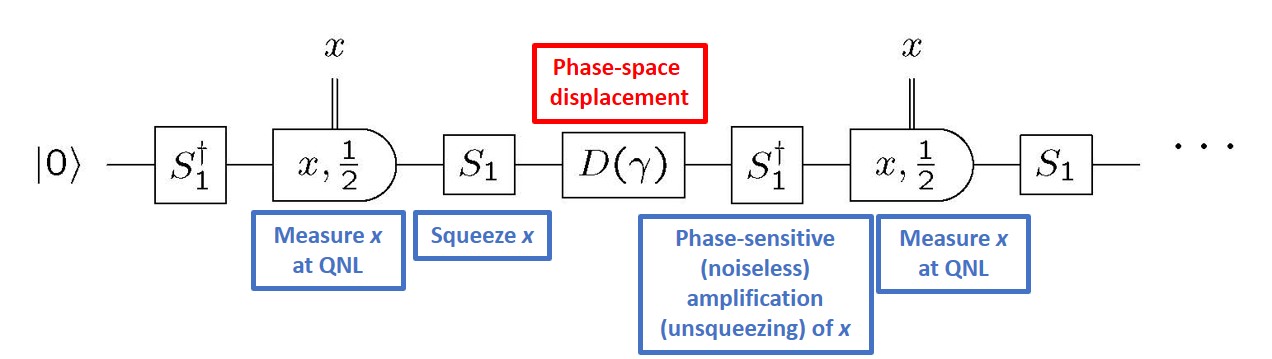}
\caption{Circuit for BAE measurement of position-quadrature displacements with resolution~$\sqrt2\sigma$.  In this circuit the high-resolution position-quadrature measurements of Fig.~\ref{fig:BAE}(a) are replaced with their equivalent from Fig.~\ref{fig:BAEtoSqueeze}(a), i.e., measurements at the QNL conjugated by squeeze operators.  The squeeze operator $S_1$ after a measurement squeezes the noise in~$x$, the signal $\gamma_1$ is imposed on top of this squeezed noise, and the second squeezer $S_1^\dagger$ noiselessly amplifies both the signal and the noise.  The amplified signal $\gamma_1e^r$ then stands out above the vacuum-level noise associated with the measurements.  The discussion in the text shows that the heterodyne measurements of Fig.~\ref{fig:BAEtoSqueeze}(c) can be substituted for the QNL-resolution position-quadrature measurements without any loss of sensitivity in estimating the displacement $\gamma_1$.}
\label{fig:BAESqueeze}
\end{figure}

The circuit equivalence in Fig.~\ref{fig:BAEtoSqueeze}(a) can be used to convert the BAE measurement circuit of Fig.~\ref{fig:BAE}(a) to an equivalent form, shown in Fig.~\ref{fig:BAESqueeze}, which uses active squeezing and quadrature measurements at the QNL in place of the high-resolution BAE measurements of Fig.~\ref{fig:BAE}(a).  It is easy to see why the circuit of Fig.~\ref{fig:BAESqueeze} works.  In the Heisenberg picture, the quadrature components evolve during the interval between measurements according to
\begin{align}\label{eq:heisenbergxp}
\begin{split}
&\hat x\mathop{\longrightarrow}_{\rm{squeeze}}^{S_1}
\hat x\,e^{-r}\mathop{\longrightarrow}_{\rm{displace}}^{D(\gamma)}
\hat x\,e^{-r}+\gamma_1\mathop{\longrightarrow}_{\rm{unsqueeze}}^{S_1^\dagger}
\hat x+\gamma_1 e^r\,,\\
&\hat p\mathop{\longrightarrow}_{\rm{unsqueeze}}^{S_1}
\hat p\,e^r\mathop{\longrightarrow}_{\rm{displace}}^{D(\gamma)}
\hat p\,e^r+\gamma_2\mathop{\longrightarrow}_{\rm{squeeze}}^{S_1^\dagger}
\hat p+\gamma_2 e^{-r}\,.
\end{split}
\end{align}
The position-quadrature displacement, smaller than the vacuum level, is imposed on top of squeezed noise, and the second squeezer acts as a noiseless, phase-sensitive amplifier that amplifies the signal so that it can be detected above the vacuum-level noise from the measurements; indeed, the overall effect on the position quadrature is to displace it by a signal amplified by the factor $e^r$.  For the momentum quadrature, just the opposite occurs, and the momentum-quadrature displacement is de-amplified below the vacuum noise.

The circuit in Fig.~\ref{fig:BAESqueeze}, has been implemented by Burd {\it et al.}~\cite{Burd2019a} to detect one quadrature of the motion of a mechanical oscillator that is a single ${}^{25}{\rm Mg}^+$ ion oscillating in an ion trap; the authors stress that the technique relies on the phase-sensitive de-amplification and amplification of squeezing as a substitute for high-resolution measurements.  Something very much like the circuit in Fig.~\ref{fig:BAESqueeze} has been used by Malnou~{\it et al.}~\cite{Malnou2019a} to demonstrate an advantage to using active squeezing of a microwave mode in the situation where dissipation dominates the measurements.  A squeeze-displace-unsqueeze proposal that is essentially identical to Fig.~\ref{fig:BAESqueeze} has been proposed for application to the collective spin of a large number of two-level atoms, with spin squeezing (one-axis twisting) in place of quadrature squeezing and collective rotations in place of phase-space displacements~\cite{EDavis2016a}.

In the next subsection, I build on the circuit of Fig.~\ref{fig:BAESqueeze} to detect both quadrature components of the displacement $\gamma$ by the simple expedient of having two modes coupled to the same classical force and doing BAE measurements of $x$ on one mode and of $p$ on the other.  The resulting two-mode circuit can be transformed to an SU(1,1) interferometer.

Before getting to that, however, I digress briefly to note that the vacuum-resolution homodyne measurements we have considered up till now can be replaced, without any loss of resolution, by heterodyne measurements.  Heterodyne measurements are represented by Kraus operators, $\sqrt{dx\,dp/2\pi}\,|\alpha\rangle\langle\alpha|$, that project onto coherent states; they describe simultaneous, quantum-limited measurements of both quadrature components~\cite{Arthurs1965a,Braunstein1991a}.    I label the quadrature outcomes by $(\alpha_1,\alpha_2)=(x,p)$, where $\alpha=(\alpha_1+i\alpha_2)/\sqrt2=(x+ip)/\sqrt2$.  The corresponding POVM elements are $dx\,dp\,|\alpha\rangle\langle\alpha|/2\pi$.  Squeezing a coherent state $|\alpha\rangle$ yields a squeezed state $|\alpha',r\rangle$,
\begin{align}
S_1|\alpha\rangle
=S_1D(\alpha)|0\rangle
=D(\alpha')S_1|0\rangle
=D(\alpha')|0,r\rangle
=|\alpha',r\rangle\,,
\end{align}
where $|0,r\rangle$ is a squeezed vacuum state, and $|\alpha',r\rangle$ is a displaced squeezed state with mean complex amplitude
\begin{align}
\alpha'=\frac{1}{\sqrt2}\big(xe^{-r}+ipe^{r}\big)=\frac{1}{\sqrt2}(x'+ip')\,.
\end{align}
These results mean that conjugating a heterodyne measurement with squeeze operators is equivalent to measuring in the squeezed basis $|\alpha',r\rangle$, an equivalence depicted in quantum circuits in Fig.~\ref{fig:BAEtoSqueeze}(c).  Measuring in this squeezed basis is the sort of BAE measurement originally contemplated by Hollenhorst~\cite{Hollenhorst1979a}.  It is as good a BAE measurement of the position quadrature as the homodyne measurements considered up till now---and maybe better because the measurement of the $p$ quadrature, poor resolution though it has, keeps the back action onto $p$ from increasing without bound and leading to states infinitely squeezed in $x$.  Moreover, measuring in the squeezed basis is certainly easier to analyze, as I now show, because the measurement outcomes determine the post-measurement state.

For this brief analysis, let us focus on heterodyne measurements conjugated by squeezing.  The $n$th heterodyne measurement, with outcome $\alpha_n$, leaves the mode, after application of the following squeeze operator, in the squeezed state $S_1|\alpha_n\rangle=|\alpha'_n,r\rangle$.  The displacement operator $D(\gamma_{n+1})$ and the subsequent unsqueezing by $S_1^\dagger$ leave the state $S_1^\dagger D(\gamma_{n+1})S_1|\alpha_n\rangle$ just before the $(n+1)$th measurement.  The outcomes of the $(n+1)$th heterodyne measurement, $\alpha_{n+1}=(x_{n+1}+ip_{n+1})/\sqrt2$, are thus drawn from the Gaussian distribution,
\begin{align}
\begin{split}
\frac{1}{2\pi}\big|\langle\alpha_{n+1}|S_1^\dagger D(\gamma_{n+1})S_1|\alpha_n\rangle\big|^2
&=\frac{1}{2\pi}\bigg|\Big\langle\alpha_{n+1}\Big|D\Big(\frac{1}{\sqrt2}(\gamma_{1,n+1}e^r+i\gamma_{2,n+1}e^{-r})\Big)\Big|\alpha_n\Big\rangle\bigg|^2\\
&=\frac{1}{\sqrt{2\pi}}e^{-(x_{n+1}-x_n-\gamma_{1,n+1}e^r)^2/2}\frac{1}{\sqrt{2\pi}}e^{-(p_{n+1}-p_n-\gamma_{2,n+1}e^{-r})^2/2}\,,
\end{split}
\end{align}
so introducing two zero-mean, unit-variance Gaussian random processes, one can write
\begin{align}
x_{n+1}&=x_n+\gamma_{1,n+1}e^r+W_{x,n+1}\,,\\
p_{n+1}&=p_n+\gamma_{2,n+1}e^{-r}+W_{p,n+1}\,.
\end{align}
The estimator for the position-quadrature displacement,
\begin{align}
\gamma_{1,n+1}^{\rm est}=(x_{n+1}-x_n)e^{-r}=\gamma_{1,n+1}+e^{-r}W_{x,n+1}\,,
\end{align}
is unbiased, i.e., $\langle\gamma_{1,n+1}^{\rm est}\rangle=\gamma_{1,n+1}$, with the variance setting the resolution,
\begin{align}
\big\langle(\Delta\gamma_{1,n+1}^{\rm est})^2\big\rangle=e^{-2r}=2\sigma^2\,.
\end{align}
This is the same as for BAE homodyne measurements, although here the factor of 2 comes from the extra half quantum of noise in simultaneous measurements of both quadratures in heterodyne measurements~\cite{Arthurs1965a,Braunstein1991a}.  An astute reader might realize that this analysis of squeezed heterodyne measurements is a rewrite of the Heisenberg-picture arrows of Eqs.~(\ref{eq:heisenbergxp}); the Heisenberg-picture arrows provide a complete analysis for heterodyne measurements because the post-measurement state is determined by the measurement outcomes.

\begin{figure}
\centering
\includegraphics[width=0.75\columnwidth]{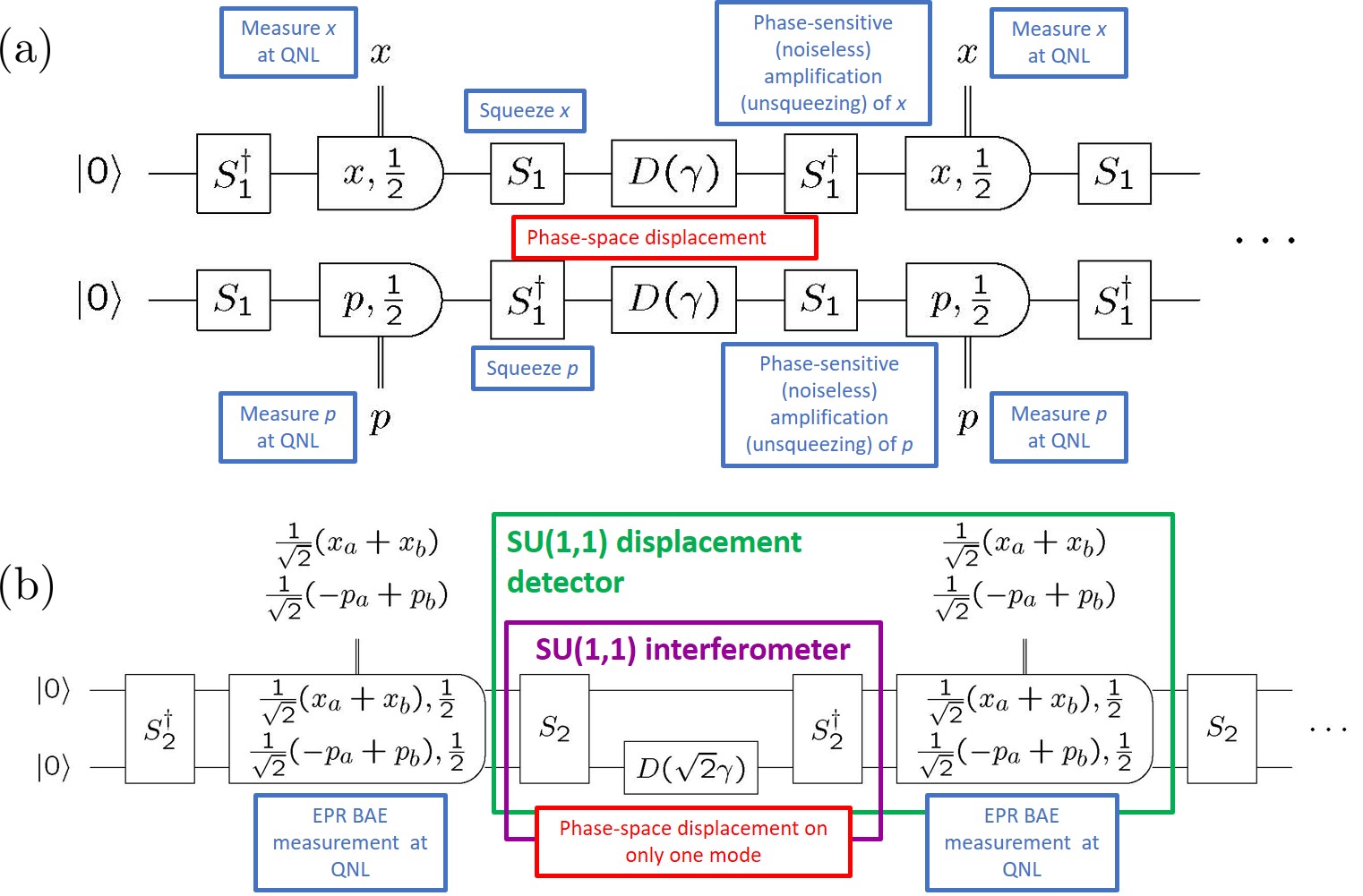}
\caption{(a)~Measuring both displacement quadratures on two modes subjected to the same displacement.  On the top mode, one measures the position quadrature~$x$, and on the bottom mode, the momentum quadrature~$p$.  The squeezing enhancements are done oppositely on the two modes so as to give subvacuum resolution for $x$ on the top mode and for $p$ on the bottom mode.  (b)~Equivalent circuit, obtained by using the three equivalences displayed in Fig.~\ref{fig:CircuitIdentities}.  What emerges in this equivalent circuit is an SU(1,1) interferometer in the middle, with the displacement signal on mode~$b$ only, and the interferometer surrounded by measurements at QNL resolution of the (commuting) EPR variables $(\hat x_a+\hat x_b)/\sqrt2$ and $(-\hat p_a+\hat  p_b)/\sqrt2$.  The initial two-mode squeezer squeezes the noise in both EPR variables, the displacement signal is imposed on top of this squeezed noise, and the second squeezer noiselessly amplifies the signal as it also amplifies the noise to the vacuum level, ready for the QNL-level measurement of the EPR variables.  The emphasis on noiseless linear amplification and de-amplification and measurement of linear observables prompts reframing of the SU(1,1) interferometer, plus the measurement, as an SU(1,1) displacement detector.  There are two ways, without loss of sensitivity, to avoid measuring joint variables in~(b).  First, one can convert the EPR measurement back to a measurement of quadrature components on separate modes by using the equivalence of Fig.~\ref{fig:CircuitIdentities}(c); what this does is to surround the two-mode squeezers with beamsplitters, thus putting them inside an SU(2) interferometer.  Second, one can replace the QNL-level quadrature measurements in~(a) with heterodyne measurements, in which case the beamsplitter transformation to~(b) leaves heterodyne measurements with (formally) relabeled outcomes.}
\label{fig:BothQuadraturestoSU11}
\end{figure}

The analysis so far confirms that the sensitivity to quadrature displacements is limited only by the available level of squeezing or high-sensitivity BAE measurements.  The practical limit on sensititivity comes from including dissipation and finite temperature, which together limit the (dimensionless) resolution to $\agt\sqrt{k_BT/\hbar\omega}\sqrt{\omega\tau/Q}=\sqrt{k_BT\tau/\hbar Q}$, where $\omega$ is the modal frequency and $\tau$ is the time between measurements.

\subsection{Measurement of both components of quadrature displacement}
\label{sec:BAEbothquadratures}

This section considers measurements of both quadrature displacements.  The obvious way to accomplish this is to have two persistent modes subjected to the same classical force and to measure $x$ on one and $p$ on the other~\cite{Hollenhorst1979a,Thorne1979a,Caves1980b,Braginsky1980a,Caves1983a}.  In the case of axion detection, for example, the discussion in the Introduction indicates that for two identical microwave cavities to see the same displacement, they need to be well within about a thousand electromagnetic wavelengths of one another.  The active-squeezing version of the two circuits for measuring both quadrature displacements is shown in Fig.~\ref{fig:BothQuadraturestoSU11}(a).  The notable feature of this circuit is that the squeezing of the persistent modes is equal and opposite to give subvacuum resolution for $x$ measurements on one mode and subvacuum resolution for $p$ measurements on the other.

\begin{figure}
\centering
\includegraphics[width=0.9\columnwidth]{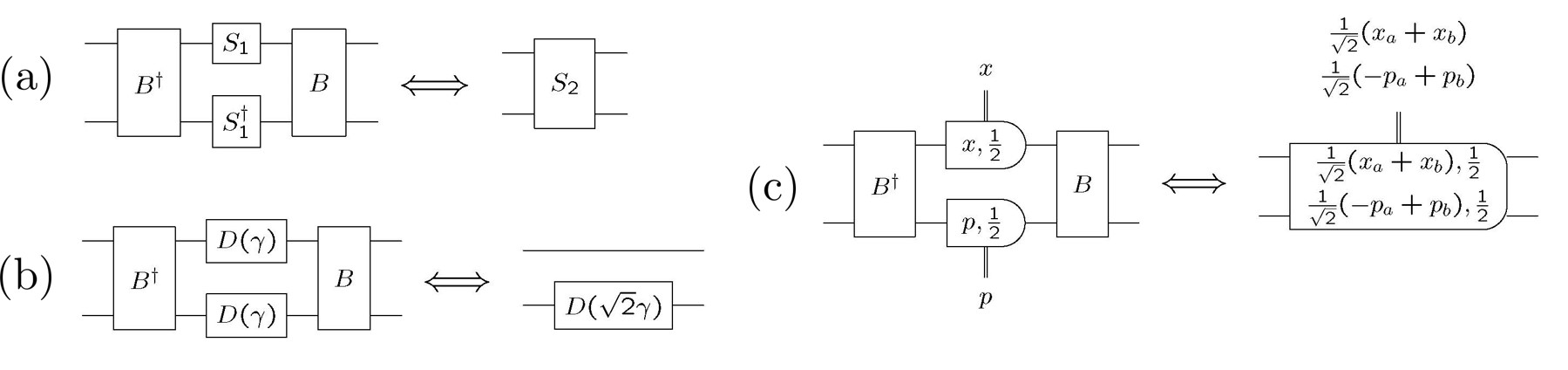}
\caption{Conjugating with the 50/50 beamsplitter $B$ of Eq.~(\ref{eq:5050bs}) performs the following transformations: (a)~equal and opposite squeezing on two modes conjugates to the two-mode squeeze operator of Eq.~(\ref{eq:S2}) $\big(BS_1\otimes S_1^\dagger B^\dagger=S_2\big)$; (b)~common-mode displacement of two modes conjugates to a displacement ($\sqrt2$ larger) on only one of the modes $\big[BD(\gamma)\otimes D(\gamma)B^\dagger=I\otimes D(\sqrt2\gamma)\big]$; (c)~quadrature measurements of $x$ on one mode and $p$ on the other conjugate to measurement of (joint) EPR variables of the two modes $\big[B(\hat x_a+i\hat p_b)B^\dagger=(\hat x_a+\hat x_b)/\sqrt2+i(-\hat p_a+\hat p_b)/\sqrt2\big]$.  In~(c), if one substitutes heterodyne measurements for the QNL-level quadrature measurements, then the beamsplitter transformation leaves heterodyne measurements with relabeled outcomes on the right.}
\label{fig:CircuitIdentities}
\end{figure}

Converting the circuit of Fig.~\ref{fig:BothQuadraturestoSU11}(a) to an SU(1,1) interferometer involves inserting $B^\dagger B=I$ between all the elements and then employing the three circuit identities of Fig.~\ref{fig:CircuitIdentities}.   The beamsplitter unitary here is the 50-50 beamsplitter of Fig.~\ref{fig:interferometry}:
\begin{align}\label{eq:5050bs}
B=e^{(\hat a\hat b^\dagger-\hat a^\dagger\hat b)\pi/4}=e^{-iJ_2\pi/2}\,.
\end{align}
It transforms the modal annihilation operators according to
\begin{align}\label{eq:BStransformab}
\begin{split}
B^\dagger\hat a B&=\frac{1}{\sqrt2}(\hat a-\hat b)\,,\\
B^\dagger\hat b B&=\frac{1}{\sqrt2}(\hat a+\hat b)\,.
\end{split}
\end{align}
The key consequence of this transformation is the well-known fact, represented in Fig.~\ref{fig:CircuitIdentities}(a), that $B$ conjugates equal and opposite squeezing of two modes to the two-mode squeeze operator~(\ref{eq:S2}):
\begin{align}
S_1(r)\otimes S_1(-r)=S_1(r)\otimes S_1^\dagger(r)=B^\dagger S_2(r)B\,.
\end{align}
In addition, it is easy to see that the common-mode displacement $\gamma$ conjugates under $B$ to a displacement of the $b$ mode [Fig.~\ref{fig:CircuitIdentities}(b)]:
\begin{align}
D(\gamma)\otimes D(\gamma)=B^\dagger I\otimes D(\sqrt2\gamma)B\,.
\end{align}

For the transformation of the quadrature measurements to measurement of the (joint) EPR variables, depicted in Fig.~\ref{fig:CircuitIdentities}(c), it is productive to introduce EPR variables as they are encoded in the (Hermitian) real and imaginary parts of complex operators:
\begin{align}\label{eq:EPR}
\begin{split}
\hat a^\dagger+\hat b&=\frac{1}{\sqrt2}(\hat x_a+\hat x_b)+i\frac{1}{\sqrt2}(-\hat p_a+\hat p_b)\,,\\
\hat a-\hat b^\dagger&=\frac{1}{\sqrt2}(\hat x_a-\hat x_b)+\frac{i}{\sqrt2}(\hat p_a+\hat p_b)\,.
\end{split}
\end{align}
Then the beamsplitter transformations~(\ref{eq:BStransformab}) can be profitably rewritten as
\begin{align}
\begin{split}
\hat x_a+i\hat p_b&=B^\dagger(\hat a^\dagger+\hat b)B\,,\\
-\hat x_b+i\hat p_a&=B^\dagger(\hat a-\hat b^\dagger)B\,.
\end{split}
\end{align}
The two EPR variables making up $\hat a^\dagger+\hat b$ transform to $\hat x_a$ and $\hat p_b$ and thus commute; likewise, the two EPR variables making up $\hat a-\hat b^\dagger$ transform to $-\hat x_b$ and $\hat p_a$ and thus commute.

The three circuit identities transform the independent-modes circuit of Fig.~\ref{fig:BothQuadraturestoSU11}(a) to the mode-coupled circuit of Fig.~\ref{fig:BothQuadraturestoSU11}(b), which has an SU(1,1) interferometer at its heart.  Because of the equivalence, one knows that the SU(1,1) circuit can determine both quadrature components of $\gamma$ with resolution $\sqrt2\sigma$, but it helps to examine the SU(1,1) circuit directly to understand how this works.  For that purpose, notice that the two-mode squeeze operator conjugates modal annihilation operators according to
\begin{align}\label{eq:S2ab}
\begin{split}
S_2^\dagger\hat a S_2&=\hat a\cosh r-\hat b^\dagger\sinh r\,,\\
S_2^\dagger\hat b S_2&=\hat b\cosh r-\hat a^\dagger\sinh r\,,
\end{split}
\end{align}
transformations that are more helpfully written in terms of the EPR observables~(\ref{eq:EPR}),
\begin{align}\label{eq:S2EPR1}
S_2^\dagger(\hat a^\dagger+\hat b)S_2&=(\hat a^\dagger+\hat b)e^{-r}\,,\\
S_2^\dagger(\hat a-\hat b^\dagger)S_2&=(\hat a-\hat b^\dagger)e^r\,.
\label{eq:S2EPR2}
\end{align}
Using these relations, one can write Heisenberg-picture arrow diagrams like those in Eqs.~(\ref{eq:heisenbergxp}),
\begin{align}\label{eq:heisenberg2}
\begin{split}
&\hat a^\dagger+\hat b\hspace{1pt}\mathop{\longrightarrow}_{\rm{squeeze}}^{S_2}\hspace{1pt}
(\hat a^\dagger+\hat b)e^{-r}\hspace{3pt}\mathop{\longrightarrow}_{\rm{displace}}^{I\otimes D(\sqrt2\gamma)}\hspace{3pt}
(\hat a^\dagger+\hat b)e^{-r}+\sqrt2\,\gamma\hspace{2pt}\mathop{\longrightarrow}_{\rm{unsqueeze}}^{S_2^\dagger}\hspace{2pt}
\hat a^\dagger+\hat b+\sqrt2\,\gamma\,e^r\,,\\
&\hat a-\hat b^\dagger\hspace{2pt}\mathop{\longrightarrow}_{\rm{unsqueeze}}^{S_2}\hspace{2pt}
(\hat a-\hat b^\dagger)e^r\hspace{3pt}\mathop{\longrightarrow}_{\rm{displace}}^{I\otimes D(\sqrt2\gamma)}\hspace{3pt}
(\hat a-\hat b^\dagger)e^r-\sqrt2\,\gamma^*\mathop{\longrightarrow}_{\rm{squeeze}}^{S_2^\dagger}
\hat a-\hat b^\dagger-\sqrt2\,\gamma^*e^{-r}\,.
\end{split}
\end{align}
Not surprisingly, the pairs of EPR variables do the same thing as the quadrature components do in the circuit of Fig.~\ref{fig:BAESqueeze}.  The EPR variables in $\hat a^\dagger+\hat b$ undergo noiseless de-amplification, acquisition of the displacement signal, which is below the vacuum level, and then noiseless amplification so that the displacement signal stands out above vacuum-level noise.  The use of noiseless amplification and de-amplification and of linear detection and the absence of any apparent role for interference prompts rethinking the ``interferometer'' in SU(1,1) interferometer; I suggest that this device, including the measurements, is more suitably called an {\it SU(1,1) displacement detector}.

An interesting feature of the circuit in Fig.~\ref{fig:BothQuadraturestoSU11}(b) is that one needs only one mode coupled to the displacement signal; one can measure both quadrature components of the displacement on that mode by measuring two of the joint EPR variables, which commute and are displaced by the two quadrature components of $\sqrt2\gamma$.  For instance, in axion detection, one cavity could be immersed in a magnetic field and thus coupled to the axion field, while the other cavity, located anywhere, is outside the magnetic field and thus not coupled to the axion field.  Indeed, the circuit of Fig.~\ref{fig:BothQuadraturestoSU11}(b) has been proposed and analyzed within the context of axion detection and modeling of microwave devices by~\cite{HZheng2016au}.

\begin{figure}
\centering
\includegraphics[width=0.7\columnwidth]{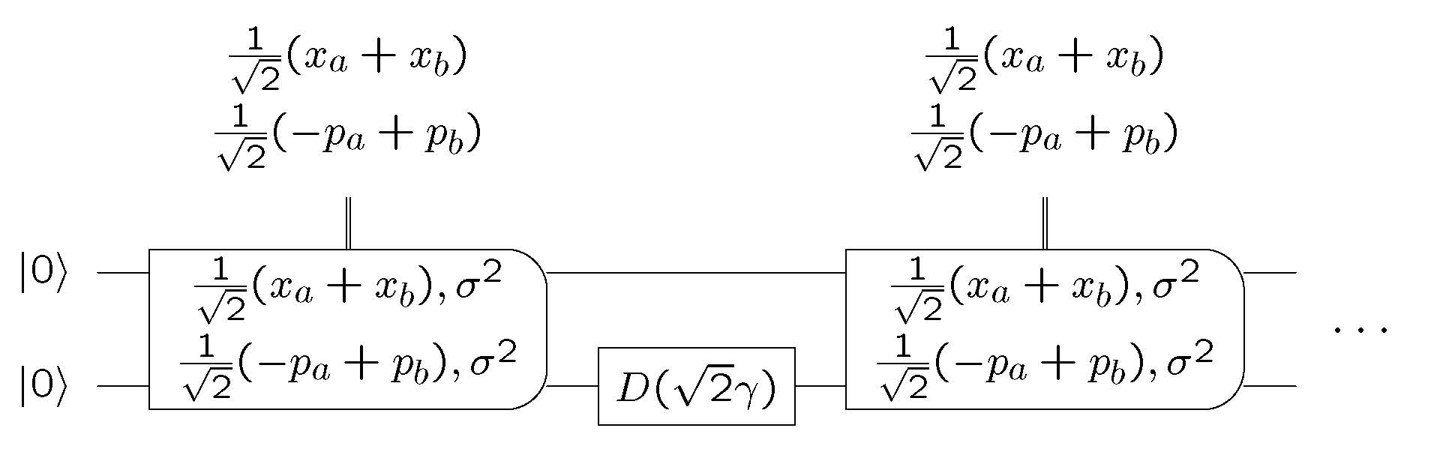}
\caption{Back to BAE measurements and no active squeezing.  The vacuum-resolution EPR measurements of Fig.~\ref{fig:BothQuadraturestoSU11}(b), enhanced by two-mode squeeze operators, have been replaced by EPR measurements with resolution~$\sigma=e^r/\sqrt2$, to create an equivalent circuit that relies only on high-resolution measurements, instead of using active squeezing.  The equivalence relies on the conjugation~(\ref{eq:S2EPR1}).}
\label{fig:EPRmeasurement}
\end{figure}

\begin{figure}
\centering
\includegraphics[width=0.9\columnwidth]{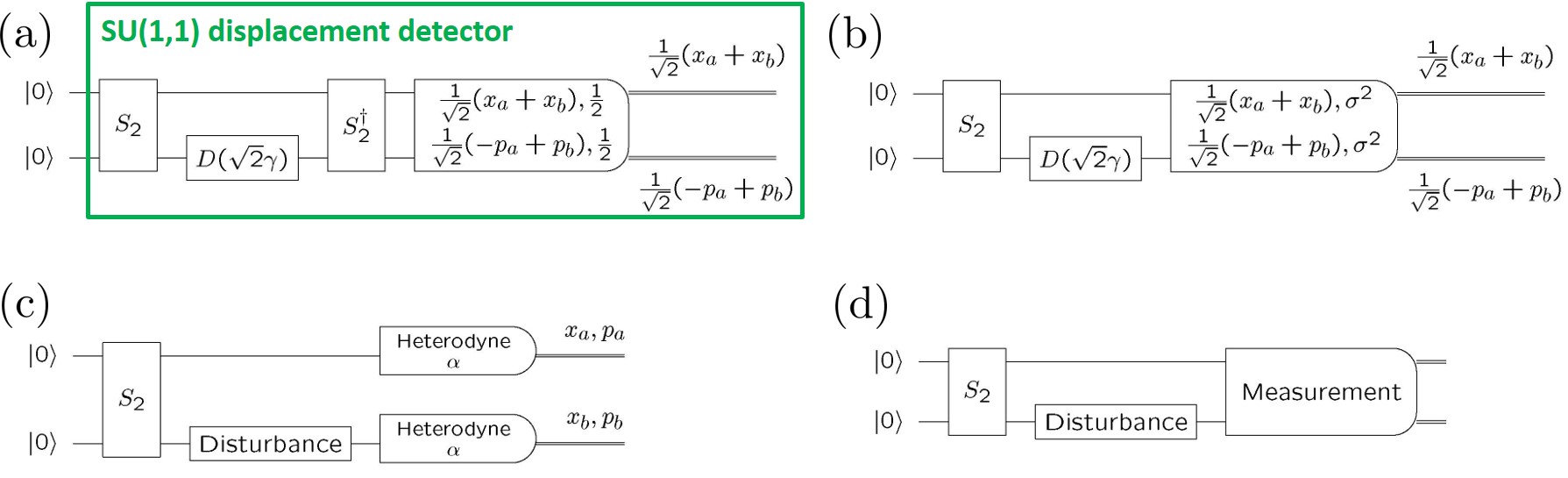}
\caption{(a)~Single trial of an SU(1,1) displacement detector for detecting a persistent displacement signal $D(\sqrt2\gamma)$ on mode~$b$.  In this circuit, the emphasis is still on making high-resolution measurements.  To do so, one makes measurements of the EPR variables at the \hbox{QNL}, or one could substitute heterodyne measurements of the two modes.  (b)~Circuit equivalent to~(a), in which the second two-mode squeezer is omitted in favor of making high-resolution measurements of the EPR variables.  The omission of the second squeezer {\it truncates\/} the SU(1,1) device and removes the last vestige of the idea that it runs on interference.  (c)~Circuit for detecting a disturbance of mode~$b$ by monitoring via heterodyne measurements all four quadrature components of the two modes.  (d)~General form of circuit for detecting a disturbance of mode~$b$ by making any kind of measurement on the two itinerant modes.  In going from~(a) and~(b) to~(c) and~(d), the emphasis shifts from making high-precision measurements that rely on the noiseless amplification and de-amplification provided by squeezing to detecting and characterizing over many trials a disturbance on mode~$b$ by taking advantage of the entangled correlations introduced by (even weak) two-mode squeezing.}
\label{fig:ToPerSigItMode}
\end{figure}

Another interesting point is that one can replace the vacuum-resolution quadrature measurements in the separate-modes circuit of Fig.~\ref{fig:BothQuadraturestoSU11}(a) with heterodyne measurements without loss of sensitivity.  Then, when one transforms to the coupled-modes circuit of Fig.~\ref{fig:BothQuadraturestoSU11}(b), one finds that the beamsplitter transforms heterodyne measurements to heterodyne measurements, but with the outcomes relabeled.  The relevant transformation of coherent-state projectors follows from the transformation~(\ref{eq:BStransformab}) of modal annihilation operators:
\begin{align}
B|\alpha\rangle\langle\alpha|\otimes|\beta\rangle\langle\beta|B^\dagger
=\Big|\frac{1}{\sqrt2}(\alpha-\beta)\Big\rangle\Big\langle\frac{1}{\sqrt2}(\alpha-\beta)\Big|
\otimes\Big|\frac{1}{\sqrt2}(\alpha+\beta)\Big\rangle\Big\langle\frac{1}{\sqrt2}(\alpha+\beta)\Big|\,.
\end{align}
The relabeling amounts to the statement that measuring all four quadrature components of the two modes is equivalent to measuring all four EPR variables.  What all this means that in Figs.~\ref{fig:BothQuadraturestoSU11}(b) and Fig.~\ref{fig:ToPerSigItMode}(a), one can substitute heterodyne measurements for the QNL-level quadrature measurements without loss of resolution.

The discussion of linear force detection in this section started with a purely measurement-based, back-action-evading protocol [Fig.~\ref{fig:BAE}(a)] that employed no active squeezing.  We can come full circle now by converting the active-squeezing circuit of Fig.~\ref{fig:BothQuadraturestoSU11}(b) to one that uses instead sub-QNL measurements of EPR variables.  The squeezing transformation~(\ref{eq:S2EPR1}) absorbs the two-mode squeezers in Fig.~\ref{fig:BothQuadraturestoSU11}(b) into the measurements of EPR variables, while changing those measurements to have sub-QNL resolution~$\sigma=e^r/\sqrt2$.  The resulting measurement-only circuit is depicted in Fig.~\ref{fig:EPRmeasurement}.

\subsection{SU(2) vs.~SU(1,1)}

A more discursive discussion of SU(2) and SU(1,1) devices is perhaps in order.

The key feature of a standard SU(2) interferometer, such as the Mach-Zehnder interferometer in Fig.~\ref{fig:interferometry}(a), is that it measures the differential phase shift between the two arms by converting it, at the second beamsplitter, to a differential amplitude change, which can be measured by differenced photodetection.  Thinking more generally, however, a small phase shift can be regarded as a phase-space displacement that is orthogonal to the mean field.  There are four possible phase-space displacements within the arms of the interferometer, those being common-mode amplitude and phase changes and differential amplitude and phase changes.  Why does an SU(2) interferometer detect only one of these, the differential phase change?

The common-mode amplitude and phase changes are equivalent to changing the amplitude and phase of the laser source; they emerge at the output as common-mode amplitude and phase changes.  Differenced photodetection is insensitive to these changes, and this insensitivity, which extends to insensitivity to common-mode noise, is part of the point of standard interferometry.  In principle, these common-mode changes could be measured, but if one has the capability to measure them at the output, one can omit the second beamsplitter and measure them within the two arms or, better yet, omit the entire interferometer and measure amplitude and phase signals directly on the input laser light.

Differential displacements in amplitude and phase within the arms are equivalent to signal coming from the dark port, thus explaining very simply why it is the noise from the dark port, masquerading as differential signal within the arms, that is the fundamental noise in interferometry.  The two differential signals exchange roles at the second beamsplitter, the differential phase changes becoming differential amplitude changes, detectable by differenced photodetection, and the differential amplitude changes becoming differental phase changes.   This is, indeed, as noted above, what an SU(2) interferometer is all about.  One could, in principle, measure the differential phase changes at the output, but the best way to do that would be to omit the second beamsplitter and to use differenced photodetection to measure those same changes as differential amplitude changes within the~arms.

One is left back at the key feature of standard interferometry, i.e., using interference at the second beamsplitter to turn  differential phase shifts between the two arms into differential amplitude changes, which are measured by differenced square-law detection.

The precise SU(1,1) analogue of the Mach-Zehnder SU(2) in Fig.~\ref{fig:interferometry}(a) is to introduce common-mode phase shifts within the arms, i.e., the unitary operator $e^{iK_0\varphi}=e^{i(\hat a^\dagger\hat a+\hat b\hat b^\dagger)\varphi/2}$, and to do summed photodetection on the output modes, i.e., to measure $K_0=\hat a^\dagger\hat a+\hat b\hat b^\dagger$.  A quick calculation shows, however, that with vacuum inputs, this only produces a signal at second order in~$\varphi$.

Far more productive is to follow the course taken in this paper and to consider detection of phase-space displacements.  The general situation is to impose phase-space displacements on both modes, i.e., to consider the displacement operator $D(\gamma_a)\otimes D(\gamma_b)$, so that the overall unitary for the device is
\begin{align}
S_2^\dagger D(\gamma_a)\otimes D(\gamma_b)S_2\,.
\end{align}
The Heisenberg-picture arrow diagram for this overall unitary, generalizing Eqs.~(\ref{eq:heisenberg2}), is
\begin{align}\label{eq:heisenberg22}
\begin{split}
&\hat a^\dagger+\hat b\hspace{1pt}\mathop{\longrightarrow}_{\rm{squeeze}}^{S_2}\hspace{1pt}
(\hat a^\dagger+\hat b)e^{-r}\hspace{3pt}\mathop{\longrightarrow}_{\rm{displace}}^{D(\gamma_a)\otimes D(\gamma_b)}\hspace{3pt}
(\hat a^\dagger+\hat b)e^{-r}+\gamma_a^*+\gamma_b\hspace{2pt}\mathop{\longrightarrow}_{\rm{unsqueeze}}^{S_2^\dagger}\hspace{2pt}
\hat a^\dagger+\hat b+(\gamma_a^*+\gamma_b)e^r\,,\\
&\hat a-\hat b^\dagger\hspace{2pt}\mathop{\longrightarrow}_{\rm{unsqueeze}}^{S_2}\hspace{2pt}
(\hat a-\hat b^\dagger)e^r\hspace{3pt}\mathop{\longrightarrow}_{\rm{displace}}^{D(\gamma_a)\otimes D(\gamma_b)}\hspace{3pt}
(\hat a-\hat b^\dagger)e^r+\gamma_a-\gamma_b^*\mathop{\longrightarrow}_{\rm{squeeze}}^{S_2^\dagger}
\hat a-\hat b^\dagger+(\gamma_a-\gamma_b^*)e^{-r}\,.
\end{split}
\end{align}
This is exactly what one would expect from our previous considerations: two EPR displacements,
\begin{align}
\gamma_a^*+\gamma_b=\frac{1}{\sqrt2}(\gamma_{a,1}+\gamma_{b,1})+i\frac{1}{\sqrt2}(-\gamma_{a,2}+\gamma_{b,2})\,,
\end{align}
are amplified so they can be detected above vacuum noise, and the other two EPR displacements,
\begin{align}
\gamma_a-\gamma_b^*=\frac{1}{\sqrt2}(\gamma_{a,1}-\gamma_{b,1})+i\frac{1}{\sqrt2}(\gamma_{a,2}+\gamma_{b,2})\,,
\end{align}
are de-amplified.

What one is able to measure with subvacuum resolution are common-mode position-quadrature displacements and differential momentum-quadrature displacements.  This accounts for the perhaps puzzling displacement signal depicted in Fig.~\ref{fig:interferometry}(b),
\begin{align}
D(\gamma^*)\otimes D(\gamma)=e^{\gamma^*(\hat a^\dagger-\hat b)-\gamma(\hat a-\hat b^\dagger)}\,,
\end{align}
which, since $\gamma_a=\gamma^*$ and $\gamma_b=\gamma$, deposits signal only in the EPR variables that can be detected with enhanced resolution, i.e.,  $\gamma_a^*+\gamma_b=2\gamma$ and $\gamma_a-\gamma_b^*=0$.

SU(1,1) devices live on the noiseless linear amplification and de-amplification of squeezers to empower linear measurements at the vacuum level to achieve subvacuum displacement resolution.  This discussion thus supports the renaming of SU(1,1) interferometers as {\it SU(1,1) displacement detectors}.\footnote{A mathy way of saying this is that SU(2) interferometry relies entirely on SU(2) operations, whereas SU(1,1) displacement detectors are all about the interaction of squeezing with the linear displacements and linear measurements of the Weyl-Heisenberg group.}  Perhaps SU(1,1)-enhanced displacement detector would be more descriptive, but nobody is likely to have the patience for that.

Let us now turn away from the discussion of persistent modes subjected to itinerant signals to just the opposite, a persistent signal that is examined repeatedly by itinerant modes.

\section{SU(1,1) detection of persistent signal using itinerant modes}
\label{sec:PerSigItMode}

\subsection{Changing perspective}
\label{sec:perspective}

With a persistent signal that can be consulted over and over again by successive itinerant modes, there is no reason to worry about the post-measurement state of the two modes in an SU(1,1) device.  This in mind, the first step is to modify the SU(1,1) displacement detector depicted in Fig.~\ref{fig:BothQuadraturestoSU11}(b) by deleting the quantum wires coming out of the EPR measurement [see Fig.~\ref{fig:ToPerSigItMode}(a)]; the only outputs from the measurement are the classical wires that carry the EPR outcomes.   This done, one can pull the same trick as in Fig.~\ref{fig:EPRmeasurement} by omitting the second two-mode squeezer and replacing it by EPR measurements that have subvacuum resolution [see Fig.~\ref{fig:ToPerSigItMode}(b)]; the deletion of the second two-mode squeezer is called {\it truncating\/} the SU(1,1) device~\cite{BAnderson2017a,BAnderson2017b,PGupta2018a}.   Up to this point, the devices under consideration are still aimed at high-resolution measurements in a single shot, but moving forward, one shifts emphasis.  If the objective is to detect and/or characterize a disturbance of mode~$b$ over many trials, one doesn't really need high-resolution measurements.  Instead, what one does is to use the entangled correlations introduced by a two-mode squeezer to distribute information about a disturbance on mode~$b$ over both quadratures of both modes and to measure all four quadrature components to get as much information about the disturbance as possible.  Thus an obvious strategy, illustrated in Fig.~\ref{fig:ToPerSigItMode}(c), is to characterize a disturbance on mode~$b$ from the statistics of heterodyne measurements on both modes; this strategy is especially suitable in the case of weak squeezing, where all four quadrature components carry roughly the same amount of information.  A yet more general strategy, depicted in Fig.~\ref{fig:ToPerSigItMode}(d), is to allow any sort of measurement on the two modes, whatever is found to be best for characterizing the disturbance at hand.

The remainder of this section is devoted to presenting briefly an example where one uses SU(1,1) techniques to detect and characterize a disturbance. Though the example is drawn from my own work, I emphasize that none of the material is original to this paper.  The example deals with {\it in situ\/} characterization of a passive linear optical network used for randomized boson sampling and is work done with Saleh Rahimi-Keshari and Sima Baghbanzadeh and reported fully in~\cite{RahimiKeshari2019au}, which the reader should consult for a complete exposition.

\subsection{\textit{\textbf{In situ\/}} characterization of linear optical networks in randomized boson sampling}
\label{sec:insitu}

\begin{figure}
\centering
\includegraphics[width=0.75\columnwidth]{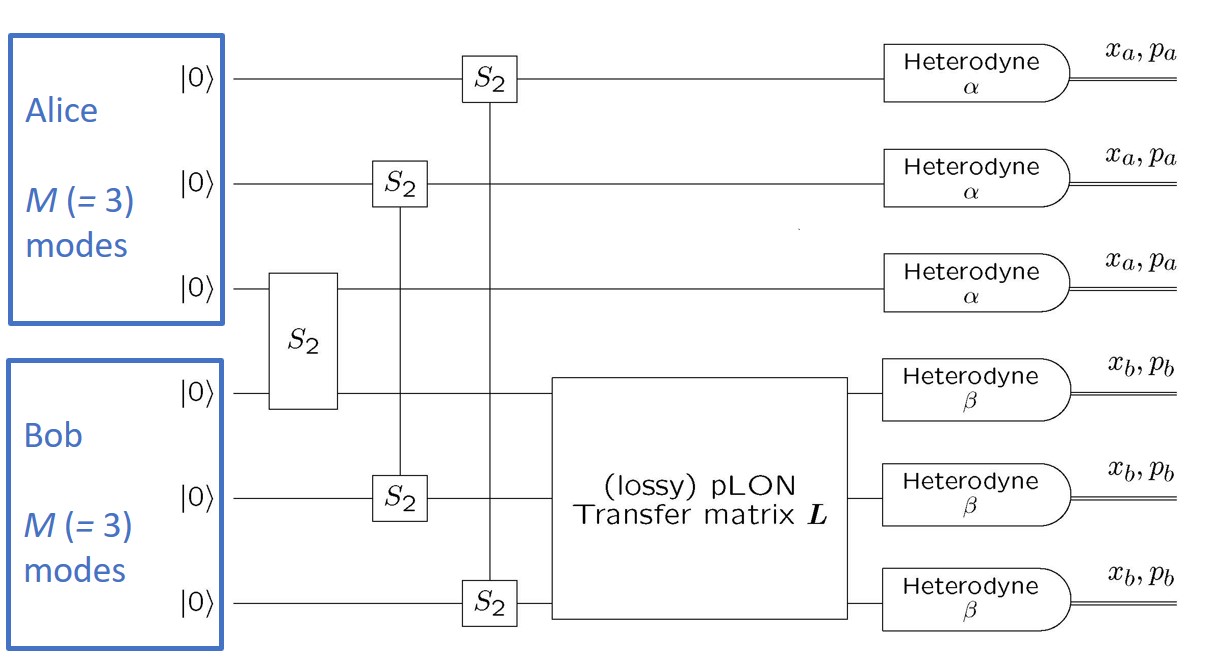}
\caption{$M$-mode version of the characterization circuit in Fig.~\ref{fig:ToPerSigItMode}(c).  Alice has $M$ modes, paired with Bob's $M$ modes.  Each pair of modes is excited into a two-mode squeezed-vacuum state.  Alice's modes go directly to heterodyne detectors; Bob's modes transit a passive linear-optical network (pLON), after which they are directed to heterodyne measurements.  As the text discusses, the transfer matrix $\bm{L}$ of the pLON can be reconstructed from the statistics of the heterodyne measurements.}
\label{fig:InSitu1}
\end{figure}

The setting now is the set of $M$ SU(1,1) devices depicted in Fig.~\ref{fig:InSitu1}.  The upper $M$ modes belong to Alice and are numbered from bottom to top; the lower $M$ modes belong to Bob and are numbered from top to bottom.  The modes are paired up by number, and each pair is excited into a two-mode squeezed-vacuum state $S_2(r)|0,0\rangle$.  The overall state of all the modes is thus
\begin{align}\label{eq:PsiAB}
\begin{split}
\ket{\Psi_{AB}}
&=\bigotimes_{j=1}^M e^{r(\hat a_j\hat b_j-\hat a_j^\dagger\hat b_j^\dagger)}|0_j,0_j\rangle\\
&=\frac{1}{\cosh^M\!r}\bigotimes_{j=1}^M e^{-\hat a_j^\dagger\hat b_j^\dagger\tanh r}|0_j,0_j\rangle\\
&=\frac{1}{\cosh^M\!r}\bigotimes_{j=1}^M\sum_{n_j=0}^\infty(-\tanh r)^{n_j}|n_j,n_j\rangle_j\,.
\end{split}
\end{align}
The second line uses the quasi-normal-ordered form of the squeeze operator~\cite{Schumaker1985a,Schumaker1986a}.  Bob's modes are input to a lossy, passive linear-optical network (pLON) that is described by an $M\times M$ transfer matrix $\bm{L}$.  The transfer matrix specifies how coherent-state amplitudes are transmitted through the pLON; i.e., a coherent state $|\beta_1,\beta_2,\ldots,\beta_M\rangle=|\bm{\beta}\rangle$, where $\bm{\beta}$ is the $M$-dimensional row vector of complex amplitudes, when input to the pLON, becomes the coherent state $|\bm{\beta}\bm{L}\rangle$ at the pLON's output.  The final step in Fig.~\ref{fig:InSitu1} is to do heterodyne measurements on all the modes, both at Alice's end and Bob's end.  The resulting circuit is the $M$-mode version of the SU(1,1) device depicted in Fig.~\ref{fig:ToPerSigItMode}(c), where the disturbance to be characterized is the \hbox{pLON}.

It is clear from the final form in Eq.~(\ref{eq:PsiAB}) that the marginal state at Alice's end and the marginal state input to Bob's pLON are identical product thermal states, with the mean number of photons in each mode being $\sinh^2\!r$.  The marginal heterodyne statistics at Alice's end are described by an isotropic Gaussian distribution whose quadrature variances are $1+\sinh^2\!r=\cosh^2\!r$.

The strict photon-number correlations between Alice's modes and Bob's modes, displayed in the final line of Eq.~(\ref{eq:PsiAB}), mean that if Alice were to count $n_j$ photons in mode~$j$, $n_j$ photons would enter the pLON from the partner mode~$j$ at Bob's end.  The case of interest to us is when the total mean number of photons in the marginal thermal states is $M\sinh^2\!r\simeq\sqrt M$; the reason for this choice is that it is the largest squeezing for which there is a reasonably good chance that a photocounting Alice would detect only vacuum or single photons, thereby preparing single photons in a random selection of about $\sqrt M$ modes at the input to Bob's \hbox{pLON}.  This is very weak squeezing as $M$ gets large.

The point of the circuit in Fig.~\ref{fig:InSitu1} is to characterize the transfer matrix $\bm{L}$.  Since the transfer matrix $\bm{L}$ transforms coherent-state amplitudes through its pLON, one can characterize the pLON by inputting a sufficient variety of coherent states and by doing heterodyne detection or photocounting at the pLON's output; procedures for selecting an appropriate set of input coherent states have been shown explicitly by~\cite{Rahimi-Keshari2013a}.  This can be achieved in the current setting by conditioning the input states to the pLON on the outcomes of Alice's heterodyne measurements.  Indeed, if Alice's heterodyne measurements yield outcomes $\begin{pmatrix}\alpha_1&\cdots&\alpha_M\end{pmatrix}=\bm{\alpha}$, the state input to the pLON, obtained by projecting onto the coherent state $|\bm{\alpha}\rangle$,
\begin{align}
\langle\bm{\alpha}|\Psi_{AB}\rangle
\propto\bigotimes_{j=1}^M e^{-\alpha_j^*\hat b_j^\dagger\tanh r}|0_j\rangle
\propto\bigotimes_{j=1}^M|{-}\alpha_j^*\tanh r\rangle=|{-}\bm{\alpha}^*\tanh r\rangle\,,
\end{align}
is the coherent state $|{-}\bm{\alpha}^*\tanh r\rangle$.  The state output from the pLON is the coherent state $|{-}\bm{\alpha}^*\bm{L}\tanh r\rangle$.  Alice's heterodyne outcomes are selected from an isotropic Gaussian distribution, so the conditional states input to Bob's pLON are certainly sufficient for characterization of $\bm{L}$.  Even though one is characterizing the pLON using the joint heterodyne statistics of both Alice and Bob, the emphasis here is on the heterodyne statistics at Bob's end, conditioned on the heterodyne outcomes at Alice's~end.

This procedure, though straightforward, is too straightforward.  One would like to turn the tables on the characterization, characterizing not using Bob's statistics conditioned on Alice's heterodyne outcomes, but rather using Alice's heterodyne statistics conditioned on Bob's outcomes.  One can see how this might be done in the following way.  Mode~$i$ at Bob's end is in a coherent state with complex amplitude~${-}\tanh r(\bm{\alpha}^*\bm{L})_i$.  The vector component here is
\begin{align}
(\bm{\alpha}^*\bm{L})_i=\sum_{j=1}^M\alpha_j^*L_{ji}=\bm{\alpha}^*\bm{L}_i\,,
\end{align}
where $\bm{L}_i=\begin{pmatrix}L_{1i}&\cdots&L_{Mi}\end{pmatrix}^T$ is the $i$th column of the transfer matrix.  Suppose Bob gets heterodyne outcome 0 for mode~$i$.  Though this doesn't mean that $0=\bm{\alpha}^*\bm{L}_i$, i.e., that $\bm{\alpha}^T$ is orthogonal to $\bm{L}_i$, it does mean that the statistics of Alice's heterodyne outcomes $\bm{\alpha}$, conditioned on Bob's getting heterodyne outcome 0 on mode~$i$, are prejudiced, just a bit for the weak squeezing we are contemplating, toward having $\bm{\alpha}^T$ orthogonal to $\bm{L}_i$.  More precisely, when conditioned on Bob's getting heterodyne outcome 0 on mode~$i$, Alice's heterodyne outcomes are drawn from a Gaussian distribution that is narrower along the direction of the complex vector $\bm{L}_i$ than in the $M$ complex directions orthogonal to $\bm{L}_i$.  From these conditional heterodyne statistics, specifically, from the second moments of the heterodyne outcomes, Alice can determine the $i$th column of the transfer matrix and, hence, all~columns.

There is an immediate problem.  Since Bob's heterodyne outcomes are drawn from a continuous distribution, the outcome for mode~$i$ is never exactly zero.  This problem is easily overcome by considering conditioning on all of Bob's heterodyne outcomes---it is clear from the discussion above that the joint heterodyne statistics are sufficient to characterize the pLON---but here we take a different tack: discretize Bob's outcomes by switching from heterodyne detection to photon counting.  Since getting heterodyne outcome 0 is equivalent to finding a mode in vacuum, all the conclusions about characterizing $\bm{L}_i$ apply to Alice's heterodyne statistics conditioned on Bob's counting no photons in mode~$i$; moreover, for weak squeezing, getting no counts is the most likely outcome for every mode at the output of the \hbox{pLON}.

\begin{figure}
\centering
\includegraphics[width=0.9\columnwidth]{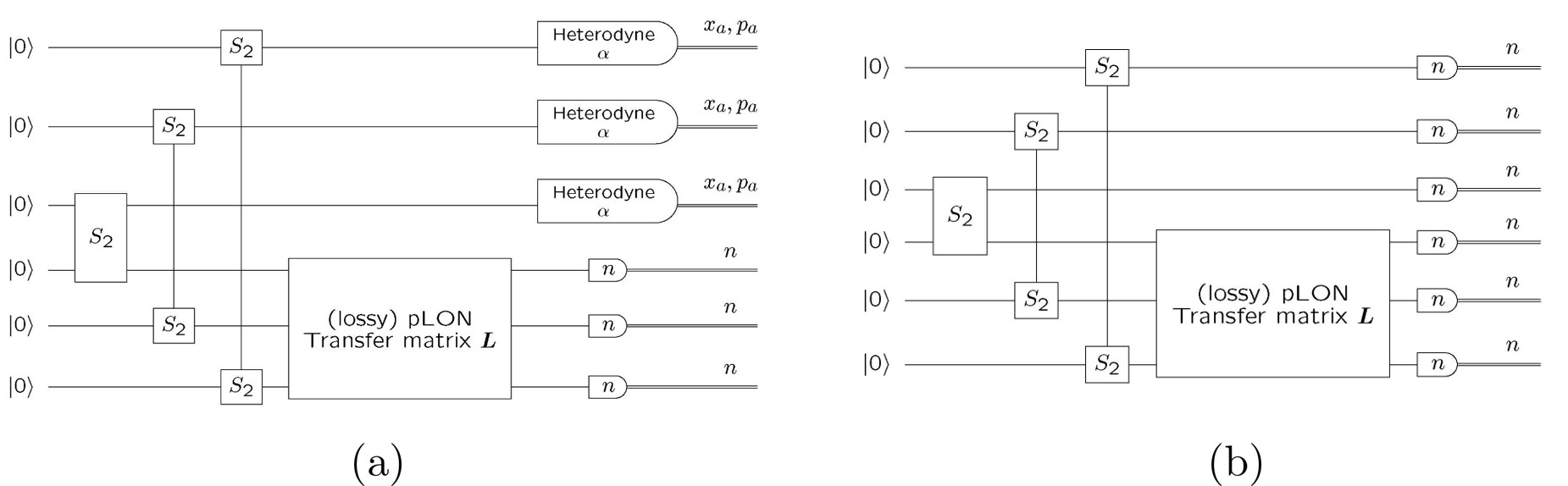}
\caption{Circuits for {\it in situ\/} characterization of a pLON used for randomized boson sampling.  (a)~Circuit for characterization runs.  Alice does heterodyne measurements, and Bob counts photons at the output of the pLON.  Alice can characterize the transfer matrix of the pLON from the heterodyne statistics conditioned on aspects of Bob's photocount record.  (b)~Circuit for randomized boson sampling.  Both Alice and Bob count photons.  Given a photocount record at Alice's end consisting of vacuum or a single photon in each mode, the input to the pLON is vacuum or a single photon into Bob's partner modes.  Bob's photocount record is an instance of the boson-sampling problem with random single-photon inputs.  The entanglement of the two-mode squeezed states is what allows Alice to decide for each run, without Bob's knowing and without changing anything at Bob's end, whether to purpose the run for characterization (by using heterodyne detection) or for boson sampling (by using photon counting).}
\label{fig:InSitu2}
\end{figure}

These considerations lead to modifying the circuit of Fig.~\ref{fig:InSitu1} to that in Fig.~\ref{fig:InSitu2}(a), where Alice makes heterodyne measurements and Bob does photocounting.  After each run of the circuit, Bob reports the photocount outcomes to Alice.  Alice gathers together the heterodyne records for which Bob reports no count in mode~$i$, estimates second moments of the modal complex amplitudes from these records, and then extracts an estimate of the $i$th column of the transfer matrix.  Proceeding in this way through all the modes at Bob's end, Alice ends up with an estimate of the entire transfer matrix.

Suppose Alice decides to count photons instead doing heterodyne measurements.  The result is the circuit of Fig.~\ref{fig:InSitu2}(b), which implements the randomized form~\cite{Lund2014a} of boson sampling~\cite{Aaronson2013a}.  When Alice counts a single photon in a mode, the companion mode at Bob's end is prepared in a single-photon state to be input to the \hbox{pLON}.  For the weak squeezing we have specified, there is a good chance that Alice counts only vacuum or a single photon in each mode; in this case, roughly $\sqrt{M}$ of Bob's modes are illuminated by a single photon, and Bob's photocounts at the output of the pLON are an instance of the boson-sampling problem, i.e., are drawn from the photocount distribution for this particular, randomly chosen set of single-photon inputs to the \hbox{pLON}.

The overall picture is now clear.  In each run of the experiment, Alice can decide to do heterodyne measurements, as in Fig.~\ref{fig:InSitu2}(a), or to do photocounting, as in Fig.~\ref{fig:InSitu2}(b); Bob does photon counting in all runs and reports the the photocount record to Alice.  The runs for which Alice does heterodyne measurements are characterization runs: Alice can extract from the heterodyne statistics of these runs, conditioned on aspects of the photocount record Bob reports, an estimate of the transfer matrix of Bob's \hbox{pLON}.  The runs for which Alice does photon counting give rise to randomized boson sampling: Alice can be confident that the photocount record Bob reports is an instance of boson sampling for single photons input to the ports of Bob's pLON corresponding to Alice's photocounts and this for the pLON described by the transfer matrix determined by the characterization runs.  The key point, which leads to calling this {\it in situ\/} characterization, is that Alice can decide which runs are used for characterization and which for randomized boson sampling without Bob's knowing and without making changes to anything at Bob's end.

The entanglement of the two-mode squeezed-vacuum states empowers the ability to do both characterization runs and boson-sampling runs without changing changing anything about the apparatus at Bob's end.  The characterization circuit in Fig.~\ref{fig:InSitu2}(a) is a multi-mode version of the general SU(1,1) device of Fig.~\ref{fig:ToPerSigItMode}(d), which is aimed at detecting and characterizing a disturbance, in this case, Bob's pLON, from the statistics of measurements on all modes.  What enables Alice to switch to boson sampling simply by doing photon counting instead of heterodyne measurements is the modal entanglement of the two-mode squeezed-vacuum states.  The characterization runs rely on the correlation between complex amplitudes at Alice's end and photon number at Bob's end, whereas the boson-sampling runs live off the (strict) correlation between Alice's photon numbers and Bob's input photon numbers.  This kind of simultaneous correlation between different physical quantities is the essence of quantum entanglement.

\section{Conclusion}

What little there is left to say can be said quickly.  The resource of SU(1,1)-generated states has played a role in quantum information since before there was a quantum information science.  The nearly 40-year-old proposal~\cite{Caves1981a} to use squeezed-vacuum light to reduce shot noise in interferometers is now topping off the billion-dollar investments in the LIGO and VIRGO gravitational-wave interferometers.  Twenty years ago, generating a two-mode squeezed-vacuum state and measuring EPR variables were the key ingredients in the first demonstration of unconditional quantum teleportation~\cite{Furusawa1998a}.  Today, ideas for modal (continuous-variable) cluster-state computation~\cite{Menicucci2006a,Menicucci2011a} live on two-mode squeezing; using this resource, experimentalists have demonstrated entanglement of large numbers of modes in one- and now two-dimensional cluster states~\cite{Asavanant2019a}.  The list goes on, but this paper does not.

Metrology is where SU(1,1) got its start in quantum information science.  In this paper, I have tried to develop a framework, based on quantum circuits, for thinking about the potential of active squeezing for metrology and other applications by relating the use of active squeezing to more primitive notions of high-resolution, back-action-evading measurements, by spelling out connections between different ways of using the resource of active squeezing, and by indicating that exploiting to the full two-mode squeezing's quantum entanglement might allow you to do more than you set out to do.

\end{document}